\documentclass[prl,aps,a4paper,twocolumn,superscriptaddress]{revtex4-1}

\usepackage[utf8x]{inputenc}
\usepackage{amssymb}
\usepackage{amsmath}
\usepackage{graphicx}
\usepackage{bbm}
\usepackage{psfrag}
\usepackage{latexsym}
\usepackage{hyperref}
\usepackage{color}

\usepackage{amsfonts}
\usepackage{amsmath}
\allowdisplaybreaks[4]        
\usepackage{amssymb}
\usepackage{euscript}     
\usepackage{color}  
\usepackage{xcolor}       
\usepackage{tensor}        
\usepackage{amsthm}
\usepackage{graphicx}
\usepackage{subfigure}
\usepackage{bbm}
\usepackage[header,title,page,titletoc]{appendix}

\makeatletter 
\def\@eqnnum{{\normalsize \normalcolor (\theequation)}} 
\makeatother

\renewcommand{\(}{\left(}
\renewcommand{\)}{\right)}
\renewcommand{\[}{\left[}
\renewcommand{\]}{\right]}

\newcommand{\cO}{{\cal O}} 
\newcommand{\bra}[1]{{\left\langle{#1}\right\vert}}
\newcommand{\ket}[1]{{\left\vert{#1}\right\rangle}}
\newcommand{\del}{\partial}

\newcommand{\be}{\begin{equation}}
\newcommand{\ee}{\end{equation}}
\newcommand{\bea}{\begin{eqnarray}}
\newcommand{\eea}{\end{eqnarray}}
\newcommand{\beq}{\begin{equation}}
\newcommand{\eeq}{\end{equation}}
\newcommand{\beqa}{\begin{eqnarray}}
\newcommand{\eeqa}{\end{eqnarray}}
\newcommand{\beqar}{\begin{eqnarray*}}
	\newcommand{\eeqar}{\end{eqnarray*}}

\newcommand{\eg}{{\it e.g.,}\ }
\newcommand{\ie}{{\it i.e.,}\ }
\newcommand{\reef}[1]{(\ref{#1})}

\newcommand{\mt}[1]{\textrm{\tiny #1}}

\newcommand{\mC}{\mathcal{C}}
\newcommand{\mD}{\mathcal{D}}
\newcommand{\mO}{\mathcal{O}}

\newcommand{\veps}{\varepsilon}
\def\S{\Sigma}

\newcommand{\ca}{{\cal C}_\mt{A}}

\newcommand{\cev}[1]{\reflectbox{\ensuremath{\vec{ \reflectbox{\ensuremath{#1}}}}}}
\newcommand{\s}{\sigma}

\newcommand{\GN}{G_\mt{N}}

\newcommand{\PsR}{\Psi_\mt{R}}
\newcommand{\PsT}{\Psi_\mt{T}}

\newcommand{\llangle}{{\langle\! \langle\,}}
\newcommand{\rrangle}{{\,\rangle\! \rangle}}

\begin{document}

\author{Alice Bernamonti}\email{abernamonti@perimeterinstitute.ca}

\affiliation{\it Perimeter Institute for Theoretical Physics, Waterloo, Ontario N2L 2Y5, Canada}
\affiliation{\it Dipartimento di Fisica, Universit\`a di Firenze and INFN Sezione di Firenze, Via G. Sansone 1, 50019 Sesto Fiorentino, Italy}

\author{Federico Galli}\email{fgalli@perimeterinstitute.ca}
\affiliation{\it Perimeter Institute for Theoretical Physics,  Waterloo, Ontario N2L 2Y5, Canada}

\author{Juan Hernandez}\email{jhernandez@perimeterinstitute.ca}
\affiliation{\it Perimeter Institute for Theoretical Physics, Waterloo, Ontario N2L 2Y5, Canada}
\affiliation{\it Department of Physics $\&$ Astronomy, University of Waterloo, Waterloo, ON N2L 3G1, Canada}

\author{Robert C. Myers}\email{rmyers@perimeterinstitute.ca}
\affiliation{\it Perimeter Institute for Theoretical Physics, Waterloo, Ontario N2L 2Y5, Canada}

\author{Shan-Ming Ruan}\email{sruan@perimeterinstitute.ca}
\affiliation{\it Perimeter Institute for Theoretical Physics, Waterloo, Ontario N2L 2Y5, Canada}
\affiliation{\it Department of Physics $\&$ Astronomy, University of Waterloo, Waterloo, ON N2L 3G1, Canada}

\author{Joan Sim\'on} \email{j.simon@ed.ac.uk}
\affiliation{\it School of Mathematics and Maxwell Institute for Mathematical Sciences,\\
	University of Edinburgh, Edinburgh EH9 3FD, UK}

\title{The First Law of Complexity}

\begin{abstract}
We investigate the variation of holographic complexity for two nearby target states. Based on Nielsen's geometric approach, we find the variation only depends on the end point of the optimal trajectory, a result which we designate the first law of complexity. As an example, we examine the complexity=action conjecture when the AdS vacuum is perturbed by a scalar field excitation, which corresponds to a coherent state. Remarkably, the gravitational contributions completely cancel and the final variation reduces to a boundary term coming entirely from the scalar field action. Hence the null boundary of Wheeler-DeWitt patch appears to act like the ``end of the quantum circuit".
\end{abstract}

\maketitle

\noindent \emph{1.~Introduction:}
Quantum information has produced  surprising new insights into foundational questions about the AdS/CFT correspondence, \eg \cite{Ryu:2006bv,Ryu:2006ef, Hubeny:2007xt, Rangamani:2016dms, Swingle:2009bg, Myers:2010tj, Blanco:2013joa, Dong:2013qoa, Faulkner:2013ica, Almheiri:2014lwa, Pastawski:2015qua}. One fascinating concept that has recently entered this discussion is {\it quantum circuit complexity}: the size of the optimal/minimal unitary circuit or transformation $U_\mt{T}$ preparing a target state $\ket{\PsT}$ from a given reference state $\ket{\PsR}$ using a set of elementary gates  \cite{Aaronson:2016vto,johnw,Susskind:2018pmk}. There have been a number of different proposals for the gravitational observables which would be dual to the complexity of a boundary state, \eg \cite{Susskind:2014rva, Stanford:2014jda,Brown:2015bva,Brown:2015lvg,Couch:2016exn}. The focus of our discussion will be the complexity=action (CA) conjecture  \cite{Brown:2015bva,Brown:2015lvg}, which suggests
%rcm: will we actually use "CA"?? yes!
\begin{equation}\label{defineCA}
\ca(\S) =  {I_\mt{WDW}}/\pi\,. 
\end{equation}
That is, the holographic complexity of a boundary state on the time slice $\S$ should be the gravitational action evaluated on the so-called Wheeler-DeWitt (WDW) patch, defined as the domain of dependence of a bulk spatial slice anchored on $\Sigma$. One important feature motivating the study of holographic complexity is that these new gravitational observables are sensitive to the bulk physics deep in the interior of a black hole \cite{Susskind:2014rva,Susskind:2014moa}. 

Exploring the properties of the new gravitational observables and their implications for complexity in the boundary theories is now an active area of research, \eg  \cite{Susskind:2014rva,Stanford:2014jda,Susskind:2014jwa, Brown:2015bva,Brown:2015lvg,Susskind:2014moa, Susskind:2015toa,Roberts:2014isa,Lehner:2016vdi,Cai:2016xho, Couch:2016exn,Reynolds:2016rvl,Chapman:2016hwi,Carmi:2016wjl, Moosa:2017yvt,Couch:2017yil,Cai:2017sjv,Brown:2017jil, Carmi:2017jqz,Swingle:2017zcd,Flory:2017ftd,Zhao:2017isy, Abt:2017pmf,Abt:2018ywl,Alishahiha:2018tep,An:2018xhv, Fu:2018kcp,Mahapatra:2018gig,Chapman:2018dem, Chapman:2018lsv,Cano:2018aqi,Barbon:2018mxk, Susskind:2018fmx,Susskind:2018tei,Cooper:2018cmb, Numasawa:2018grg,Brown:2018kvn,Goto:2018iay, Agon:2018zso, Chapman:2018bqj,Flory:2018akz,Flory:2019kah}. A basic shortcoming of this research program is that we lack a proper understanding of circuit complexity of quantum field theories (QFTs). In particular, this prevents more than qualitative tests of the gravitational results. Hence, a second line of inquiry has become to fully develop the concept of circuit complexity for QFT states, in particular for states in a strongly coupled CFT (such as a holographic boundary theory), \eg \cite{ Belin:2018bpg, Belin:2018fxe, Bhattacharyya:2018wym, Takayanagi:2018pml, Caputa:2017urj, Caputa:2017yrh, Jeff,Chapman:2017rqy, Hackl:2018ptj,Khan:2018rzm, cohere, Bhattacharyya:2018bbv, Chapman:2018hou, Alves:2018qfv,Camargo:2018eof, Ali:2018fcz, Jiang:2018nzg, Roberts:2016hpo, Hashimoto:2017fga, Czech:2017ryf,Reynolds:2017lwq, Magan:2018nmu,Caputa:2018kdj,Balasubramanian:2018hsu, Jackiw:1984je,Shimaji:2018czt,Ali:2018aon,Liu:2019aji}. This will be essential to properly test the various holographic proposals and ultimately, to produce a derivation of one (or more) of these conjectures. Our objective here is to begin to build a concrete bridge between these two research directions. In particular, we examine variations of the target state and demonstrate a natural interpretation connecting both approaches.

Nielsen's geometric approach \cite{nielsen2006quantum, nielsen2008, Nielsen:2006} gives a framework to describe the complexity of QFT states, as illustrated for certain simple QFTs, \eg \cite{Jeff,Chapman:2017rqy,Khan:2018rzm,Hackl:2018ptj, Ali:2018fcz,Bhattacharyya:2018bbv,cohere,Alves:2018qfv,Camargo:2018eof, Chapman:2018hou,  Jiang:2018nzg}. It constructs a continuum representation of the unitary transformations 
	\begin{equation}\label{unitaries}
	U(\s) = \cev{\mathcal{P}}\, \exp \!\[ -i \int^\s_0\!\!\! d s\, H( s)\]\,, 
	\end{equation}
where $s$ parametrizes the circuit and $\cev{\mathcal{P}} $ signifies a right-to-left path ordering. The `Hamiltonian' $H(s)= \sum Y^I(s)\,\mathcal{O}_I$ is constructed from the (Hermitian) generators $\mathcal{O}_I$ of the elementary gates, and $Y^I(s)$ are control functions  specifying which gates (and how many times they) are applied at any point $s$ in the circuit.  Eq.~\reef{unitaries} actually specifies a path $U(\s)$ through the space of unitaries, or through the space of states with $\ket{\psi(\s)}=U(\s)\ket{\psi_\mt{R}}$. With  $\sigma\in[0,1]$,  the boundary conditions are 
	\begin{equation}
	U(\s=0)= \mathbbm{1}\,, \qquad  U(\s=1)= U_\mt{T}\,,
	\label{bc88}
	\end{equation}
where $\ket{\PsT}= U_\mt{T}\, \ket{\PsR}$. 

Introducing a set of coordinates $x^a$ on the space of unitaries described by eq.~\reef{unitaries} (or on the space of states $U(x^a)\ket{\PsR}$), paths are described as $x^a(s)$. Nielsen's approach \cite{nielsen2006quantum, nielsen2008, Nielsen:2006} identifies the optimal circuit producing $U_\mt{T}$  by minimizing the cost
\begin{equation}\label{def_com}
\mD\[x^a(s)\]= \int^1_0 ds ~ F \( x^a, \dot x^a \)\,,
\end{equation}
subject to the boundary conditions \reef{bc88}. The circuit complexity is then the cost for the optimal trajectory, \ie
\begin{equation}\label{complexity2}
\mC\( \ket{\PsT}\) \equiv \text{Min} \,\mD\,.
\end{equation}
As indicated in eq.~\reef{def_com}, the cost function $F$ is chosen as a particular functional of the position $x^a(s)$ and the tangent vector $\dot x^a(s)$.
Hence determining the complexity is analogous to the physical problem of identifying a particle trajectory by minimizing the action with Lagrangian $F ( x^a, \dot x^a )$ and then evaluating the on-shell action. The precise form of $F$ is not fixed, but reasonable cost functions satisfy a number of preferred features \cite{Nielsen:2006}: 1) Smoothness, 2) Positivity, 3) Triangle inequality and 4) Positive homogeneity -- see also \cite{Jeff,Hackl:2018ptj}.

In the context of holography, we do not yet have a clear picture of the reference state, nor of the gates, nor of the path. However, the target state has a sharp interpretation in the AdS/CFT framework. Indeed, we are interested in quantum states in the boundary CFT which are dual to smooth geometries in the bulk gravitational theory. Hence it is natural to examine the role of the latter in holographic complexity. In particular, we consider variations of the holographic complexity under changes of the target state and examine what information we can extract about the corresponding cost function.

\begin{figure}[t]
	\includegraphics[width=3.475in]{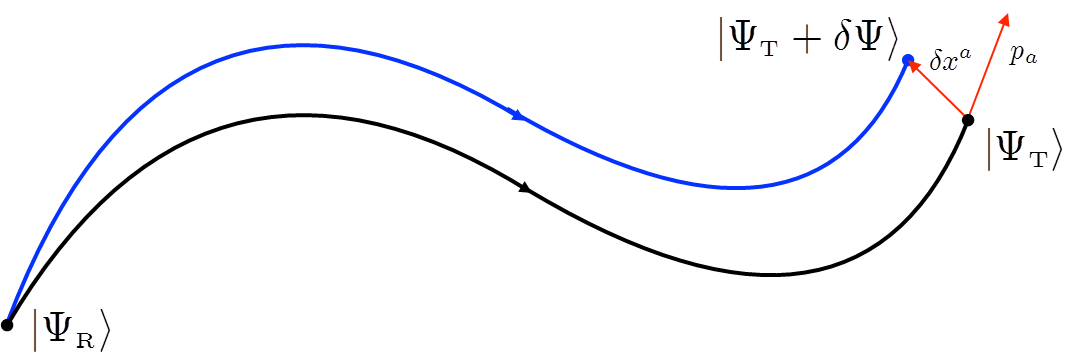}
	\caption{\label{variation} The variation of the Nielsen circuit due to a perturbation $\ket{\Psi_{\mt{T}} +\delta\Psi}$ of the target state $\ket{\Psi_{\mt{T}}}$. }
\end{figure}

\vspace{4 pt}
\noindent \emph{2.~The first law of complexity:} Using the analogy to classical mechanics, the variation of complexity \reef{complexity2} due to changing the target state with a fixed reference state, as illustrated in figure \ref{variation}, yields 
\beq
\delta\mC = p_a\, \delta x^a\big|_{s=1}\qquad{\rm with}\quad p_a =\frac{\partial F}{\partial \dot{x}^a}
\label{deltaC1}
\eeq
for any differentiable cost function $F$. The significant feature of this result, which we designate as the `{\it first law of complexity}', is that $\delta\mC$ only has contributions from the endpoint. Hence in the holographic setting, we can hope to extract information about the (variation of the) cost function in terms of bulk data describing the target state. 

If the direction along the path $p_a$ is orthogonal to the variation of the target state $\delta x^a$, 
%-- see figure \ref{variation}, 
the first order contribution \reef{deltaC1} vanishes. However, the next order variation still comes from the endpoint,
\begin{equation}
\begin{split}
  \delta\mC &=  \frac{1}{2}\,\delta p_a\, \delta x^a\big|_{s=1}\qquad {\rm with}\\
   \delta p_a &= \delta x^{b}\,\frac{\partial^2 F}{\partial x^b\partial \dot{x}^a}  +\delta \dot{x}^b\, \frac{\partial^2 F}{\partial \dot{x}^{b}\partial  \dot{x}^{a}}\,.
\end{split}
\label{deltaC2}
\end{equation}

To explore the first law of complexity in the context of holography, we consider the AdS vacuum as our original target state and the backreaction of a free bulk scalar field with a small amplitude as the perturbed target state. Evaluating the variation of eq.~\reef{defineCA} yields the change in the corresponding holographic complexity. Now this excited state can be thought of as a coherent state of the bulk scalar. In particular, the scalar field can be expressed as 
\begin{equation}\label{scalq}
\hat{\Phi}(y^\mu) = \sum%_{n} 
\(  u_{n}(y^\mu)\, {a}_{n} +  u^*_{n}(y^\mu)\, {a}_{n}{}^{\!\dagger} \) \,,
\end{equation}
where $u_{n}$ are eigenfunctions solving the Klein-Gordon equation in the AdS background. The ${a}_{n}$ and ${a}_{n}{}^{\!\dagger}$ %associated with eigenenergy $\omega_n$ 
denote the annihilation and creation operators acting on the scalar vacuum $|0\rangle$. We will assume that $y^\mu=(\rho,t,\Omega)$ denote global coordinates on the AdS background and then the sum over $n$ in eq.~\reef{scalq} includes the radial and angular quantum numbers. The excited state in which a few modes $\left\{j\right\}$
%(designated by quantum numbers $j$) 
are given a classical expectation value  can be described as a coherent state \begin{equation}\label{Def_coherent}
 \ket{\varepsilon \alpha_{j}}=  e^{\varepsilon \sum
 %_{\left\{j\right\} } 
D(\alpha_{j})}  \ket{0} \ \ {\rm with}\ \ 
 D(\alpha_{j})= {\alpha_{j} a_{j}{}^{\!\dagger}-\alpha^{\ast}_{j}a_{j}}\,,
 %= \prod_j e^{-\frac 12 |\alpha_j|^2} e^{\alpha_j a^\dagger_j}  \ket{0}\,,
\end{equation}
where we have included a small parameter $\varepsilon\ll1$ to control the overall amplitude of the scalar field
\begin{equation}
\bra{\varepsilon \alpha_j} \hat{\Phi}  \ket{\varepsilon \alpha_j}=\varepsilon
\sum
%_{\left\{j\right\}}
\!\big(\alpha_j\,u_j+\alpha_j^\ast\,u^\ast_j \big)\equiv \veps\,\Phi_{\mt{cl}}\,.
\label{class}
\end{equation}

The reader will notice that our description of the perturbed state has been entirely in terms of the bulk theory while the aim of holographic complexity is to compute the complexity of states in the boundary theory. However, the AdS/CFT correspondence simply states that the bulk and boundary theories provide alternative descriptions of the same quantum states, \ie  the vacuum state $|0\rangle$ and the Hilbert space spanned with the ${a}_{n}$ and ${a}_{n}{}^{\!\dagger}$. Hence, while the details of the description change in terms of the boundary CFT, the perturbed states in eq.~\reef{Def_coherent} are still the same coherent states in the boundary theory. 
Further, $\delta\ca$ is the variation of the complexity between these coherent states and the vacuum in the boundary theory. The bulk description of these states  lends itself to the holographic calculations %necessary 
for $\delta\ca$. 

\vspace{4 pt}
\noindent \emph{3.~Holographic Framework:} Our example begins with a four-dimensional bulk theory, Einstein gravity coupled to a negative cosmological constant and a free massless scalar field,
\beq
I_{\mt{bulk}} = \frac1{16\pi\GN} \int d^{4} y \sqrt{-g} \bigg[{\cal R}+ \frac{6}{L^2}-
\frac1{2} \nabla^\mu \Phi \nabla_\mu \Phi  \bigg]\,.
\label{Bact} 
\eeq
%For simplicity, we focus on a four-dimensional bulk (or a $d=3$ boundary CFT).  
Its vacuum $\text{AdS}_{4}$ solution is
%metric in global coordinates, %$y^\mu=(t,\rho,\theta,\phi)$, 
\begin{equation} 
ds^2_{\mt{AdS}}= \frac{L^2}{\cos^2 \rho}\( -dt^2+d\rho^2 + \sin^2\!\rho\ d\Omega_2^{2}\)\,,
%\, (d\theta^2+\sin^2\theta\,d\phi^2)
\label{vac}
\end{equation}
where $L$ denotes the AdS radius of curvature and the (dimensionless) radial coordinate 
$\rho$ runs from 0 to $\pi/2$, at the asymptotic boundary.

We perturb the vacuum by turning on the scalar in a coherent state \reef{Def_coherent}, and the classical field $\Phi_\mt{cl}$ then backreacts on the spacetime geometry. Our calculation  makes a perturbative expansion in $\varepsilon$ controlling the amplitude of the scalar in eq.~\reef{class}. 
While the full set of modes carry quantum numbers $n=(j,\ell,m)$, we focus on spherically symmetric configurations $\ell=m=0$. The scalar equation in the AdS background reduces to
\begin{equation}
%
%0 &= \frac{1}{\tan^2 \rho}\partial_\rho \( \tan^2\rho \partial_\rho\Phi \)  -\frac{M^2L^2}{\cos ^2\rho}\Phi-\partial_t^2\Phi \\
0=
 \partial^2_\rho\Phi + \frac{2}{\sin \rho\cos \rho} \partial_\rho\Phi -\partial_t^2\Phi\,.
\end{equation}
The corresponding eigenfunctions in eq.~\reef{scalq} are 
\beq\label{eigenmode}
u_j(t,\rho) =   e_j(\rho)\ e^{-i \omega_j t}
\eeq
with frequency $\omega_j =3 +2j$ and radial profile \footnote{ See Supplemental Material-A for details where the references \cite{foot01,Avis:1977yn,Burgess:1984ti,Cotabreveescu:1999em,Cotabreveescu:1999em,Fitzpatrick:2011jn,kaplan2013lectures,Terashima:2017gmc,Ammon:2015wua,ElShowk:2011ag,Fitzpatrick:2011jn,kaplan2013lectures,Terashima:2017gmc,Terashima:2017gmc,long-paper,kaplan2013lectures,BottaCantcheff:2015sav, Marolf:2017kvq, BottaCantcheff:2019apr} are included. \label{SuppMa}} 
  \begin{equation}
 \begin{split}
 e_j(\rho) \equiv & \,4\, (-)^j\,\sqrt{\frac{\GN\,(j+2)(j+1)}{\pi\,L^2\,(j+\frac32)}} \\
  &\quad\times\ \cos^3\!\rho\ {}_2 F_1\!\(-j,j+3;\ \frac{3}{2};\,\sin^2\! \rho\)\, .
 \end{split}
 \label{mode}
 \end{equation}
For simplicity, we will focus on real parameters $\alpha_j$ in eq.~\reef{Def_coherent} which then yields 
\begin{equation}
\Phi_{\mt{cl}}(t,\rho)=2
\sum%_{\left\{j\right\}} 
 \alpha_j\,e_j(\rho)\,\cos(\omega_j t) \,.\label{class2}
\end{equation}

Next we turn to the backreaction, where we follow closely the analysis in \cite{Bizon:2011gg, Buchel:2012uh,Buchel:2013uba}. For our spherically symmetric configurations \reef{class2}, we use  the metric ansatz
\begin{equation}\label{metric}
ds^2 = \frac{L^2}{\cos^2 \rho}\left(- a\,e^{-2d}dt^2 + \frac{d\rho^2}{a} + \sin^2\!\rho\ d\Omega_2^{2}\right) \,,
\end{equation}
where $a(t,\rho)$, $d(t,\rho)$ describe the metric perturbation.
Working in the small amplitude expansion, we write
\begin{equation}
\begin{split}\label{perturb}
a(t,\rho) &= 1+ \varepsilon^2\,a_2(t,\rho)  +\mO( \varepsilon^4) \,, \\
e^{d(t,\rho)}&=1+  \varepsilon^2\,d_2(t,\rho)  +\mO( \varepsilon^4)\,,\\
\end{split}
\end{equation}
and at ${\cal O}(\veps^2)$, Einstein's equations reduce to three linear first-order differential equations
\begin{equation}\label{eq:osc-linear}
\begin{split}
\partial_\rho a_2& + \frac{3-2\cos^2\! \rho}{\cos\rho\,\sin \rho}\, a_2 = \partial_\rho d_2  \,,\\
\partial_\rho d_2 &=- \frac14\, \sin \rho\,\cos \rho\ \Big( (\partial_\rho\Phi_{\mt{cl}})^2 + (\partial_t\Phi_{\mt{cl}})^2 \Big)\,, \\
\partial_t a_2 &= - \frac12\, \sin \rho\,\cos \rho\, \partial_\rho\Phi_{\mt{cl}}\, \partial_t \Phi_{\mt{cl}} \,,
\end{split}
\end{equation}
with the third being redundant. 
%For a given choice of $\alpha_j$ and 
Imposing the regularity condition $a_2(t,\rho=0)=0$ and the boundary condition $d_2(t,\rho=\pi/2)=0$, the perturbations $a_2(t,\rho)$ and $d_2(t,\rho)$ can be integrated in terms of $\Phi_{\mt{cl}}(t,\rho)$  \cite{Buchel:2012uh,Buchel:2013uba}.

\vspace{4 pt}
\noindent \emph{4.~Holographic Complexity:} The 
variation of holographic complexity evaluated to second order in $\veps$ by the CA conjecture \reef{defineCA} splits into two classes of contributions:
\begin{equation}
\delta\mC_A(\Sigma) = \frac{1}{\pi}\left(
\delta I_{\mt{WDW}}+ \delta I_{\delta\mt{WDW}}\right)\,,
\label{varC}
\end{equation}
where $ \delta I_{\mt{WDW}}$ is the variation due to the change in the background fields within the original WDW patch, while $\delta I_{\delta\text{WDW}}$ is the variation due to the change in the shape (\ie the position of the boundary) 
of the WDW patch. 

First, we must recall that as well as the bulk terms appearing in eq.~\reef{Bact}, the gravitational action includes a number of surface terms  \cite{Lehner:2016vdi,foot04}. In the present case  \footnote{ See Supplemental Material-B for more details where the references \cite{foot02,PhysRevLett.28.1082, PhysRevD.15.2752,foot03} are included.}, only two will be relevant in evaluating the variation \reef{varC}: The first is the null surface term
\begin{equation}\label{kappa}
  I_{\kappa} =  \frac{1}{8\pi \GN}\int_{\partial \mt{WDW}} \!\!\!\!\!\!ds\,d^2\Omega\,\sqrt{\gamma}\,\kappa\,,
\end{equation}
where $\gamma$ is the metric determinant on the boundary of the WDW patch. $\kappa$ describes the failure of the coordinate $s$ along the null boundary to be affine, \ie $k^\mu\nabla_\mu k_\nu = \kappa\,k_\nu$ where $k_\mu dx^\mu$  is the outward-directed null normal. The second term
\begin{equation}\label{ct33}
  I_{\mt{ct}} =  \frac{1}{8\pi \GN}\int_{\partial \mt{WDW}} \!\!\!\!\!\!ds\,d^2\Omega\,\sqrt{\gamma}\,\Theta\log(\ell_{\mt{ct}}\Theta)\,,
\end{equation}
ensures the action is invariant under reparametrizations of the null boundaries \cite{Lehner:2016vdi}.  Here, $\Theta=\partial_s\log \sqrt{\gamma}$ is the expansion scalar of the null generators on the boundary, and  $\ell_{\mt{ct}}$ is an arbitrary scale needed for the argument of the logarithm to be dimensionless.

If we consider the WDW patch anchored at $t=0$ in the AdS vacuum \reef{vac}, the future and past null boundaries are given by $t=t_\pm(\rho)=\pm(\pi/2 - \rho)$, and we choose the null normals as $k_\mu dx^\mu=\pm dt+d\rho$. The  boundary coordinate is implicitly defined by $\partial_s\equiv k^\mu\partial_\mu=\cos^2\rho/L^2(\mp \partial_t + \partial\rho$).

In the perturbed background, the null boundaries experience a small shift $\delta t_\pm(\rho)$, which is determined by
\beq
\partial_\rho \delta t_\pm = \pm\, \veps^2 \left.(a_2-d_2)\right|_{t=t_\pm(\rho)}\,.
\label{eq:naff}
\eeq

The variation $\delta I_{\mt{bulk}}$ contributes to $\delta I_{\delta\mt{WDW}}$ in eq.~\reef{varC} by the boundary integrals of $\delta t_\pm(\rho)$ times the bulk action evaluated for the AdS vacuum, \ie ${\cal R}+{6}/{L^2}=-6/L^2$, and to $\delta I_{\mt{WDW}}$ yielding a total derivative, which is evaluated as a surface integral on the undeformed boundaries $t=t_\pm(\rho)$. Hence the entire variation $\delta\mC_A(\Sigma)$ is given by surface integrals along the boundary of the WDW patch.

Turning to the contributions coming from eqs.~\reef{kappa} and \reef{ct33}. Above, we chose an affine boundary coordinate $s$, giving $\kappa=0$ at leading order. The simplest approach is to keep the same coordinate for the second order calculations  and, at this order, it fails to be affine. 
Hence we have a nonvanishing variation $\delta I_\kappa$ with
\begin{equation}
  \kappa = \pm  \varepsilon^2 \frac{\cos^2\rho}{L^2} \del_t (a_2-d_2)\,.
\end{equation}
The variation $\delta I_\mt{ct}$ reduces to
\begin{equation}\label{hop99}
 \delta I_{\mt{ct}} = \frac{1}{8\pi \GN}\int_{\partial \mt{WDW} } \!\!\!\!\!\!ds\,d^2\Omega\ \delta k^\mu\,\partial_\mu\sqrt{\gamma}\,.
\end{equation}
We note that this result is independent of the scale $\ell_{\mt{ct}}$.

Combining all contributions, the holographic complexity variation equals
\begin{equation}
  \delta\mC_A(\Sigma) = \frac{\delta I_{\mt{mat}}}\pi = -\frac{\veps^2}{64\pi^2\GN} \int_{\partial\mt{WDW}}\!\!\!\!\!\!ds\,d^2\Omega\sqrt{\gamma}\,\partial_s \( \Phi_{\mt{cl}}^{\,2} \)\,.
\label{cavar} 
\end{equation}
Above we are emphasizing that the sum of all gravitational contributions precisely cancels and the full variation comes entirely from the variation of the scalar field action $\delta I_{\mt{mat}}$. Given the state in eq.~\reef{Def_coherent}, this becomes
\begin{equation}\label{wow1}
\delta\mC_A(\Sigma) =\frac{\varepsilon^2}{\pi^2} \sum\limits_{j_1,j_2}  \alpha_{j_1}\, \alpha_{j_2}\, C_{j_1,j_2}\ ,
\end{equation}
with the coefficients given by 
\begin{widetext}
\begin{equation}\label{wow2}
C_{j_1,j_2}= \sqrt{\frac{(j_1+\frac32)(j_2+\frac32)}{\left(j_1+1\right) \left(j_1+2\right)\left(j_2+1\right) \left(j_2+2\right)}} 
\times\, \bigg( H_{j_1+\frac{1}{2}}+H_{j_1+\frac{3}{2}} +H_{j_2+\frac{1}{2}}+H_{j_2+\frac{3}{2}}  
-H_{j_1+j_2+\frac{5}{2}}-H_{j_1-j_2-\frac{1}{2}}-2+4\log 2\bigg)\,.
\end{equation}
\end{widetext}
The $H_\alpha=\partial_\alpha \log\Gamma(\alpha+1)+\gamma_0 $ are harmonic numbers (where $\gamma_0$ is Euler's constant). The reflection relation $H_{-\alpha-1}-H_{\alpha}= \pi \cot(\pi \alpha )$ guarantees  $C_{j_1,j_2} = C_{j_2,j_1}$. 

\begin{figure}[t]
	\includegraphics[width=3.475in]{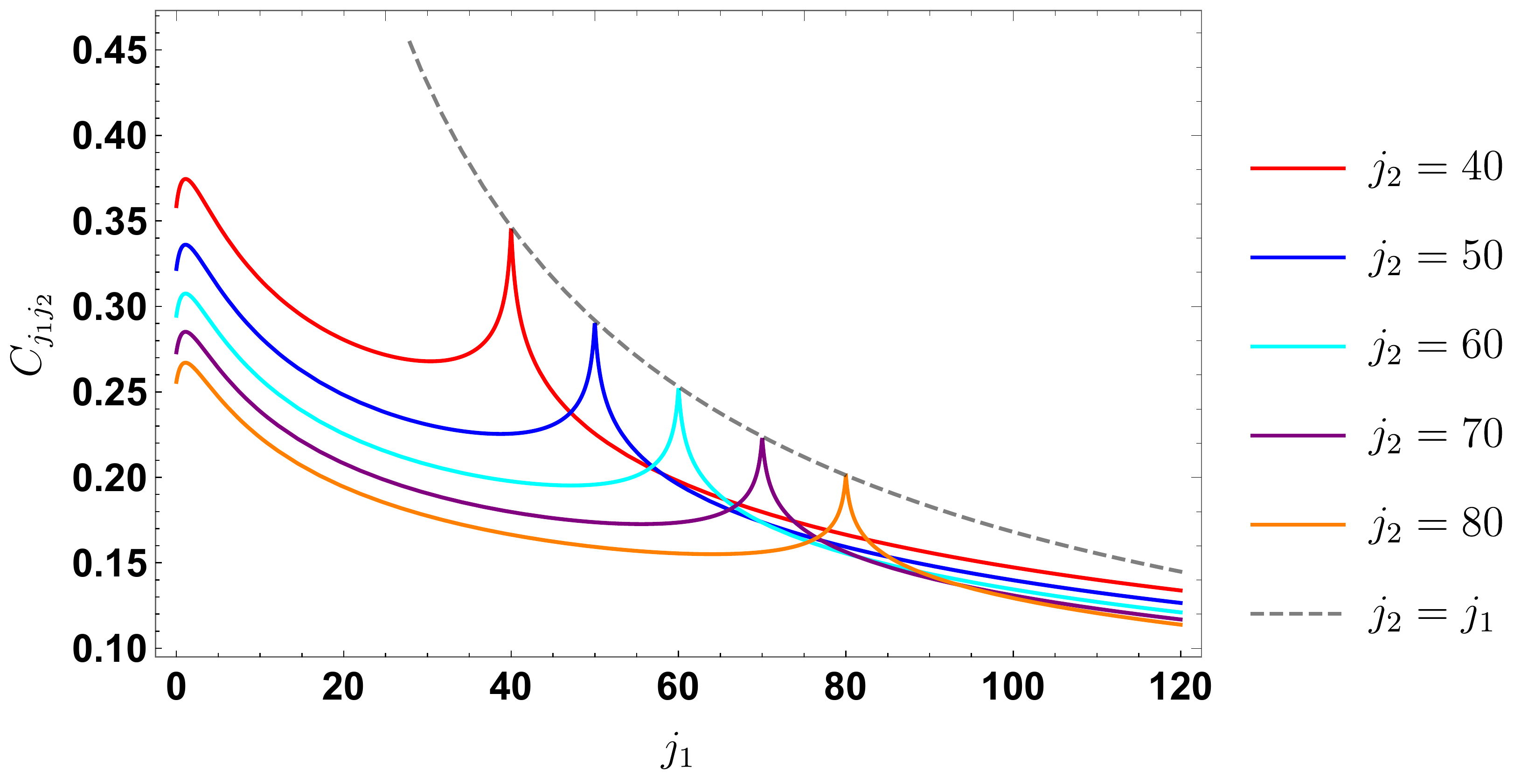}
	\caption{\label{Cj1j2} The plot for $C_{j_1,j_2}$ with fixed $j_2$. Each curve has peaks at $j_1=1$ and $j_1=j_2$. The envelope of the latter is shown with the dashed gray line. Although we draw continuous curves to help guide the eye, one should only think of $j_1$ as taking integer values, \ie $j_1=0,1,2,\cdots$.}
\end{figure}

Figure \ref{Cj1j2} shows $C_{j_1, j_2}$ as a function of $j_1$ for various values of $j_2$. We can see that these curves have two peaks, one at $j_1=1$ and the other at $j_1=j_2$. However, in both instances, the peak value decays as $j_2$ grows. In fact, near the diagonal peak one has
\beq
\lim_{j\to\infty} C_{j,j+\delta j}=3\,\frac{\log 2j}{j}+\mathcal{O}\left(\frac{1}{j}\right)
\label{bugs}
%&&+\frac{3 \gamma_0 +3 \log 2-2-H_{-\delta j-\frac{1}{2}} }{ j} +\mathcal{O}\left(\frac{\log j}{j^2}\right)\,, 
%\nonumber
\eeq
with fixed $\delta j$. This further indicates that the curves are relatively flat for large $j$ since  $\delta j$ only appears in the $\mathcal{O}(1/j)$ term. We might also note that for large $j_1$ (and fixed $j_2$), the curves are decaying as $(j_1-j_2)^{-1/2}$.

\vspace{4 pt}
\noindent \emph{5.~Discussion:} We applied the first law of complexity to examine variations of holographic complexity generated by a small scalar field excitation in  AdS. 
Our results in eqs.~\reef{wow1} and \reef{wow2} depend only on the {\it dimensionless} parameters, $\varepsilon$ and $\alpha_j$, characterizing these excitations \reef{Def_coherent}. These are coherent states of the bulk scalar, but are equally described as coherent states of the dual marginal operator (and its descendants) in the boundary theory.  Hence these parameters and $\delta\mC_A$ have a natural interpretation as boundary quantities.  
It is interesting that $\delta\mC_A$ is scale independent, in contrast to the full holographic complexity, \eg where the leading  UV divergence has the form $\mC_A \sim \log\!\(2\ell_\mt{ct} /L\)  {\rm Vol}(\Sigma) / \delta^{2}$ with  $\delta$ being the short-distance cutoff \cite{Carmi:2016wjl,Vad2}.

Given that our final result \reef{wow1} is second order, \ie $\delta\mC_A\sim\varepsilon^2\alpha^2$, the first order variation $p_a\, \delta x^a$ in eq.~\reef{deltaC1} must vanish. That is, the cost function appropriate for the CA conjecture defines a geometry where the coherent state directions are orthogonal to the direction along the circuit preparing the CFT vacuum. Instead, the leading variation of the holographic complexity in this example takes the form given in eq.~\reef{deltaC2}. 

It is difficult to interpret this result without further assumptions. For example, let us assume that the cost function has the simple form $F=g_{ab}\dot x^a \dot x^b$, known as the $\kappa=2$ measure \cite{cohere}. Then the vanishing of the first order variation indicates that at the end-point of the circuit, the off-diagonal components of the metric $g_{ab}$ between the coherent state and vacuum preparation directions vanish. If this vanishing holds in the vicinity of the end-point, \ie it also holds for the first derivative of the metric, then the remaining variation of the complexity takes the form: $\delta\mC= g_{ab}\,\delta x^a\delta\dot x^b$.  In this scenario, comparing to eq.~\reef{wow1},  the coefficients can be interpreted directly as metric components on the corresponding space, \ie $C_{j_1,j_2}\sim g_{j_1j_2}$.

We can compare our holographic results to the variation of the complexity for a free massless scalar field in a fixed AdS geometry \reef{vac} by evaluating the circuit complexity of the vacuum state and the coherent states \reef{Def_coherent} following \cite{Jeff,cohere}. In this set-up, the reference state is an unentangled state of local scalar field degrees of freedom in the AdS space, while in holography, it corresponds to an unentangled state of quantum gravity degrees of freedom (and so presumably there is no spacetime). The circuit complexity in the QFT depends on the choice of the cost function (see \cite{cohere,long-paper} for further details), however, a characteristic result for the $\kappa=2$ measure is 
\begin{equation}\label{smalla}
\delta \mC_{\kappa=2} = \sum \frac{ 2\, \varepsilon^2 \alpha^2_n}{\mu^2x_0^2\,(\omega_n/R\mu-1)}\,\log\!\Big(\frac{\omega_n}{R\mu}
\Big)   \, , %+ {\cal O}(\varepsilon^4 \alpha^4)\,,
\end{equation} 
where $\omega_n=3+2j+\ell$ is the eigenfrequency for modes with $n=(j,\ell,m)$, $\mu$ is the frequency characterizing the reference state, $x_0$ is a scale appearing in the definition of the gates producing the coherent state \cite{cohere} and $R$ is an additional length scale introduced to produce a dimensionful time in the metric \reef{vac}. This QFT variation is second order, \ie the coherent state directions are orthogonal to the direction of the circuit preparing the QFT vacuum, as in our holographic framework. Its large radial quantum number limit, \ie  $\omega_n\sim 2j$ matches the large $j$ limit of the holographic result given in  eq.~\reef{bugs}. In contrast to the holographic result, all the coherent state directions are mutually orthogonal due to the orthogonality of the scalar modes \reef{eigenmode} making eq.~\reef{smalla} diagonal (with $j=j_1=j_2$). Furthermore, the absence of scales in the holographic result would require the QFT scales to be dependent, \eg $\mu x_0\sim 1 \sim \mu R$.

There is an important assumption in our derivation of the first law of complexity. When the complexity \reef{complexity2} is described as the minimal cost of circuits preparing the desired target state, we mean the global minimum over all possible circuits. When we perturb the target state, we assume that the circuit which globally minimizes the cost remains close to the original optimal circuit, \ie the family of globally minimizing circuits is continuous in the amplitude of the perturbation. While one can imagine examples where this is not the case (\eg geodesics between ``nearly'' conjugate points on a sphere), our expectation is that this assumption is valid for the example studied here. In particular, it is  explicitly seen in the QFT complexity calculations \cite{cohere}. Of course, it would also be interesting to identify situations (in either QFT or holography) where our assumption does not hold.

Returning to the holographic calculations, we recall the inclusion of the counterterm \reef{ct33} was essential for the cancellation of the gravitational contributions to the action variation leading to $\delta\mC_A$ entirely determined by the scalar field contributions. This feature may add to the essential role of this boundary term for the CA proposal \cite{Chapman:2018dem,Vad2}, despite not being necessary to have a well-defined variational principle for the gravitational action \cite{Lehner:2016vdi}. It would be interesting to better understand this cancellation and how generally it applies, \eg does it hold beyond spherical symmetry.

Irrespective of the previous cancellation, another feature of our calculations was that all of the contributions reduced to surface contributions on the boundaries of the WDW patch. In analogy to the derivation of our first law, this property essentially arises because we are considering variations of the bulk action evaluated around background on-shell configurations. In the case of the Nielsen geometry, the boundary contribution comes from the (target state) end of the circuit, \eg see Figure \ref{variation}. Hence we are led to speculate that the boundary of the WDW patch may correspond to the `end of the circuit' in the CA conjecture. This suggests a picture where the AdS spacetime is built up through adding layers of null cones. This interpretation may have connections with the surface/state correspondence of \cite{Miyaji:2015yva}.

The first law of complexity provides a new approach to investigate holographic complexity and in particular, to build a concrete bridge to standard approaches to circuit complexity. While we have provided one application of this method here, this is only a starting point. It will be straightforward to extend our calculations to other fields (\eg massive scalars or gravitons), higher spacetime dimensions, or other quantum states. The same approach can also be used to investigate the complexity=volume \cite{Susskind:2014rva, Stanford:2014jda} and complexity=spacetime volume \cite{Couch:2016exn} conjectures. 
While we initially assumed that complexity is defined by a Nielsen geometry, a similar extremization arises in the Fubini-Study approach of \cite{Chapman:2017rqy} and in the path integral optimization procedure of \cite{Bhattacharyya:2018wym, Takayanagi:2018pml, Caputa:2017urj, Caputa:2017yrh}. Hence our approach should be useful to investigate these directions as well. More generally, it provides a unified perspective with which to investigate variations of holographic complexity, \eg see \cite{Flory:2018akz,Flory:2019kah,  Belin:2018fxe, Belin:2018bpg, Bhattacharyya:2018wym}. We will explore several of these questions in \cite{long-paper}.

\vspace{8 pt}

\begin{acknowledgments}
\noindent \emph{Acknowledgments:}
It is a pleasure to thank Alex Belin, Alex Buchel, Pawel Caputa, Horacio Casini, Vincent Chen, Jordan Cotler, Jos\'e M. Figueroa-O'Farrill, Ling-Yan Hung, Javier Magan, Alex Maloney, Hugo Marrochio, James Sully, Tadashi Takayanagi and Jingxiang Wu for useful comments and conversations.  Research at Perimeter Institute is supported by the Government of Canada through the Department of Innovation, Science and Economic Development and by the Province of Ontario through the Ministry of Research \& Innovation. 
AB acknowledges support by the program ``Rita Levi Montalcini" for your young researchers and the INFN initiative GAST.  
JPH and SMR are supported in part by a Discovery Grant awarded to RCM by the Natural Sciences and Engineering Research Council of Canada. JPH is also supported by the Natural Sciences and Engineering Research Council of Canada through a NSERC PGS-D.
RCM was supported in part by research funding from the Simons Foundation through the ``It from Qubit'' Collaboration.  JS is supported by the Science and Technology Facilities Council (STFC) [grant number ST/L000458/1]. JS would also like to thank the Perimeter Institute for all their support and hospitality during the period January-June 2017 when this project started.
AB, FG and RCM thank the Galileo Galilei Institute for Theoretical Physics for hospitality and the INFN for partial support during part of this work. 

\end{acknowledgments}

\bibliography{complex}

\begin{thebibliography}{112}%
\makeatletter
\providecommand \@ifxundefined [1]{%
 \@ifx{#1\undefined}
}%
\providecommand \@ifnum [1]{%
 \ifnum #1\expandafter \@firstoftwo
 \else \expandafter \@secondoftwo
 \fi
}%
\providecommand \@ifx [1]{%
 \ifx #1\expandafter \@firstoftwo
 \else \expandafter \@secondoftwo
 \fi
}%
\providecommand \natexlab [1]{#1}%
\providecommand \enquote  [1]{``#1''}%
\providecommand \bibnamefont  [1]{#1}%
\providecommand \bibfnamefont [1]{#1}%
\providecommand \citenamefont [1]{#1}%
\providecommand \href@noop [0]{\@secondoftwo}%
\providecommand \href [0]{\begingroup \@sanitize@url \@href}%
\providecommand \@href[1]{\@@startlink{#1}\@@href}%
\providecommand \@@href[1]{\endgroup#1\@@endlink}%
\providecommand \@sanitize@url [0]{\catcode `\\12\catcode `\$12\catcode
  `\&12\catcode `\#12\catcode `\^12\catcode `\_12\catcode `\%12\relax}%
\providecommand \@@startlink[1]{}%
\providecommand \@@endlink[0]{}%
\providecommand \url  [0]{\begingroup\@sanitize@url \@url }%
\providecommand \@url [1]{\endgroup\@href {#1}{\urlprefix }}%
\providecommand \urlprefix  [0]{URL }%
\providecommand \Eprint [0]{\href }%
\providecommand \doibase [0]{http://dx.doi.org/}%
\providecommand \selectlanguage [0]{\@gobble}%
\providecommand \bibinfo  [0]{\@secondoftwo}%
\providecommand \bibfield  [0]{\@secondoftwo}%
\providecommand \translation [1]{[#1]}%
\providecommand \BibitemOpen [0]{}%
\providecommand \bibitemStop [0]{}%
\providecommand \bibitemNoStop [0]{.\EOS\space}%
\providecommand \EOS [0]{\spacefactor3000\relax}%
\providecommand \BibitemShut  [1]{\csname bibitem#1\endcsname}%
\let\auto@bib@innerbib\@empty
%</preamble>
\bibitem [{\citenamefont {Ryu}\ and\ \citenamefont
  {Takayanagi}(2006{\natexlab{a}})}]{Ryu:2006bv}%
  \BibitemOpen
  \bibfield  {author} {\bibinfo {author} {\bibfnamefont {S.}~\bibnamefont
  {Ryu}}\ and\ \bibinfo {author} {\bibfnamefont {T.}~\bibnamefont
  {Takayanagi}},\ }\href {\doibase 10.1103/PhysRevLett.96.181602} {\bibfield
  {journal} {\bibinfo  {journal} {Phys. Rev. Lett.}\ }\textbf {\bibinfo
  {volume} {96}},\ \bibinfo {pages} {181602} (\bibinfo {year}
  {2006}{\natexlab{a}})},\ \Eprint {http://arxiv.org/abs/hep-th/0603001}
  {arXiv:hep-th/0603001 [hep-th]} \BibitemShut {NoStop}%
%%CITATION = HEP-TH/0603001;%%
\bibitem [{\citenamefont {Ryu}\ and\ \citenamefont
  {Takayanagi}(2006{\natexlab{b}})}]{Ryu:2006ef}%
  \BibitemOpen
  \bibfield  {author} {\bibinfo {author} {\bibfnamefont {S.}~\bibnamefont
  {Ryu}}\ and\ \bibinfo {author} {\bibfnamefont {T.}~\bibnamefont
  {Takayanagi}},\ }\href {\doibase 10.1088/1126-6708/2006/08/045} {\bibfield
  {journal} {\bibinfo  {journal} {JHEP}\ }\textbf {\bibinfo {volume} {08}},\
  \bibinfo {pages} {045} (\bibinfo {year} {2006}{\natexlab{b}})},\ \Eprint
  {http://arxiv.org/abs/hep-th/0605073} {arXiv:hep-th/0605073 [hep-th]}
  \BibitemShut {NoStop}%
%%CITATION = HEP-TH/0605073;%%
\bibitem [{\citenamefont {Hubeny}\ \emph {et~al.}(2007)\citenamefont {Hubeny},
  \citenamefont {Rangamani},\ and\ \citenamefont {Takayanagi}}]{Hubeny:2007xt}%
  \BibitemOpen
  \bibfield  {author} {\bibinfo {author} {\bibfnamefont {V.~E.}\ \bibnamefont
  {Hubeny}}, \bibinfo {author} {\bibfnamefont {M.}~\bibnamefont {Rangamani}}, \
  and\ \bibinfo {author} {\bibfnamefont {T.}~\bibnamefont {Takayanagi}},\
  }\href {\doibase 10.1088/1126-6708/2007/07/062} {\bibfield  {journal}
  {\bibinfo  {journal} {JHEP}\ }\textbf {\bibinfo {volume} {07}},\ \bibinfo
  {pages} {062} (\bibinfo {year} {2007})},\ \Eprint
  {http://arxiv.org/abs/0705.0016} {arXiv:0705.0016 [hep-th]} \BibitemShut
  {NoStop}%
%%CITATION = ARXIV:0705.0016;%%
\bibitem [{\citenamefont {Rangamani}\ and\ \citenamefont
  {Takayanagi}(2017)}]{Rangamani:2016dms}%
  \BibitemOpen
  \bibfield  {author} {\bibinfo {author} {\bibfnamefont {M.}~\bibnamefont
  {Rangamani}}\ and\ \bibinfo {author} {\bibfnamefont {T.}~\bibnamefont
  {Takayanagi}},\ }\href {\doibase 10.1007/978-3-319-52573-0} {\bibfield
  {journal} {\bibinfo  {journal} {Lect. Notes Phys.}\ }\textbf {\bibinfo
  {volume} {931}},\ \bibinfo {pages} {pp.1} (\bibinfo {year} {2017})},\ \Eprint
  {http://arxiv.org/abs/1609.01287} {arXiv:1609.01287 [hep-th]} \BibitemShut
  {NoStop}%
%%CITATION = ARXIV:1609.01287;%%
\bibitem [{\citenamefont {Swingle}(2012)}]{Swingle:2009bg}%
  \BibitemOpen
  \bibfield  {author} {\bibinfo {author} {\bibfnamefont {B.}~\bibnamefont
  {Swingle}},\ }\href {\doibase 10.1103/PhysRevD.86.065007} {\bibfield
  {journal} {\bibinfo  {journal} {Phys. Rev.}\ }\textbf {\bibinfo {volume}
  {D86}},\ \bibinfo {pages} {065007} (\bibinfo {year} {2012})},\ \Eprint
  {http://arxiv.org/abs/0905.1317} {arXiv:0905.1317 [cond-mat.str-el]}
  \BibitemShut {NoStop}%
%%CITATION = ARXIV:0905.1317;%%
\bibitem [{\citenamefont {Myers}\ and\ \citenamefont
  {Sinha}(2011)}]{Myers:2010tj}%
  \BibitemOpen
  \bibfield  {author} {\bibinfo {author} {\bibfnamefont {R.~C.}\ \bibnamefont
  {Myers}}\ and\ \bibinfo {author} {\bibfnamefont {A.}~\bibnamefont {Sinha}},\
  }\href {\doibase 10.1007/JHEP01(2011)125} {\bibfield  {journal} {\bibinfo
  {journal} {JHEP}\ }\textbf {\bibinfo {volume} {01}},\ \bibinfo {pages} {125}
  (\bibinfo {year} {2011})},\ \Eprint {http://arxiv.org/abs/1011.5819}
  {arXiv:1011.5819 [hep-th]} \BibitemShut {NoStop}%
%%CITATION = ARXIV:1011.5819;%%
\bibitem [{\citenamefont {Blanco}\ \emph {et~al.}(2013)\citenamefont {Blanco},
  \citenamefont {Casini}, \citenamefont {Hung},\ and\ \citenamefont
  {Myers}}]{Blanco:2013joa}%
  \BibitemOpen
  \bibfield  {author} {\bibinfo {author} {\bibfnamefont {D.~D.}\ \bibnamefont
  {Blanco}}, \bibinfo {author} {\bibfnamefont {H.}~\bibnamefont {Casini}},
  \bibinfo {author} {\bibfnamefont {L.-Y.}\ \bibnamefont {Hung}}, \ and\
  \bibinfo {author} {\bibfnamefont {R.~C.}\ \bibnamefont {Myers}},\ }\href
  {\doibase 10.1007/JHEP08(2013)060} {\bibfield  {journal} {\bibinfo  {journal}
  {JHEP}\ }\textbf {\bibinfo {volume} {08}},\ \bibinfo {pages} {060} (\bibinfo
  {year} {2013})},\ \Eprint {http://arxiv.org/abs/1305.3182} {arXiv:1305.3182
  [hep-th]} \BibitemShut {NoStop}%
%%CITATION = ARXIV:1305.3182;%%
\bibitem [{\citenamefont {Dong}(2014)}]{Dong:2013qoa}%
  \BibitemOpen
  \bibfield  {author} {\bibinfo {author} {\bibfnamefont {X.}~\bibnamefont
  {Dong}},\ }\href {\doibase 10.1007/JHEP01(2014)044} {\bibfield  {journal}
  {\bibinfo  {journal} {JHEP}\ }\textbf {\bibinfo {volume} {01}},\ \bibinfo
  {pages} {044} (\bibinfo {year} {2014})},\ \Eprint
  {http://arxiv.org/abs/1310.5713} {arXiv:1310.5713 [hep-th]} \BibitemShut
  {NoStop}%
%%CITATION = ARXIV:1310.5713;%%
\bibitem [{\citenamefont {Faulkner}\ \emph {et~al.}(2014)\citenamefont
  {Faulkner}, \citenamefont {Guica}, \citenamefont {Hartman}, \citenamefont
  {Myers},\ and\ \citenamefont {Van~Raamsdonk}}]{Faulkner:2013ica}%
  \BibitemOpen
  \bibfield  {author} {\bibinfo {author} {\bibfnamefont {T.}~\bibnamefont
  {Faulkner}}, \bibinfo {author} {\bibfnamefont {M.}~\bibnamefont {Guica}},
  \bibinfo {author} {\bibfnamefont {T.}~\bibnamefont {Hartman}}, \bibinfo
  {author} {\bibfnamefont {R.~C.}\ \bibnamefont {Myers}}, \ and\ \bibinfo
  {author} {\bibfnamefont {M.}~\bibnamefont {Van~Raamsdonk}},\ }\href {\doibase
  10.1007/JHEP03(2014)051} {\bibfield  {journal} {\bibinfo  {journal} {JHEP}\
  }\textbf {\bibinfo {volume} {03}},\ \bibinfo {pages} {051} (\bibinfo {year}
  {2014})},\ \Eprint {http://arxiv.org/abs/1312.7856} {arXiv:1312.7856
  [hep-th]} \BibitemShut {NoStop}%
%%CITATION = ARXIV:1312.7856;%%
\bibitem [{\citenamefont {Almheiri}\ \emph {et~al.}(2015)\citenamefont
  {Almheiri}, \citenamefont {Dong},\ and\ \citenamefont
  {Harlow}}]{Almheiri:2014lwa}%
  \BibitemOpen
  \bibfield  {author} {\bibinfo {author} {\bibfnamefont {A.}~\bibnamefont
  {Almheiri}}, \bibinfo {author} {\bibfnamefont {X.}~\bibnamefont {Dong}}, \
  and\ \bibinfo {author} {\bibfnamefont {D.}~\bibnamefont {Harlow}},\ }\href
  {\doibase 10.1007/JHEP04(2015)163} {\bibfield  {journal} {\bibinfo  {journal}
  {JHEP}\ }\textbf {\bibinfo {volume} {04}},\ \bibinfo {pages} {163} (\bibinfo
  {year} {2015})},\ \Eprint {http://arxiv.org/abs/1411.7041} {arXiv:1411.7041
  [hep-th]} \BibitemShut {NoStop}%
%%CITATION = ARXIV:1411.7041;%%
\bibitem [{\citenamefont {Pastawski}\ \emph {et~al.}(2015)\citenamefont
  {Pastawski}, \citenamefont {Yoshida}, \citenamefont {Harlow},\ and\
  \citenamefont {Preskill}}]{Pastawski:2015qua}%
  \BibitemOpen
  \bibfield  {author} {\bibinfo {author} {\bibfnamefont {F.}~\bibnamefont
  {Pastawski}}, \bibinfo {author} {\bibfnamefont {B.}~\bibnamefont {Yoshida}},
  \bibinfo {author} {\bibfnamefont {D.}~\bibnamefont {Harlow}}, \ and\ \bibinfo
  {author} {\bibfnamefont {J.}~\bibnamefont {Preskill}},\ }\href {\doibase
  10.1007/JHEP06(2015)149} {\bibfield  {journal} {\bibinfo  {journal} {JHEP}\
  }\textbf {\bibinfo {volume} {06}},\ \bibinfo {pages} {149} (\bibinfo {year}
  {2015})},\ \Eprint {http://arxiv.org/abs/1503.06237} {arXiv:1503.06237
  [hep-th]} \BibitemShut {NoStop}%
%%CITATION = ARXIV:1503.06237;%%
\bibitem [{\citenamefont {Aaronson}(2016)}]{Aaronson:2016vto}%
  \BibitemOpen
  \bibfield  {author} {\bibinfo {author} {\bibfnamefont {S.}~\bibnamefont
  {Aaronson}}\ }(\bibinfo {year} {2016})\ \Eprint
  {http://arxiv.org/abs/1607.05256} {arXiv:1607.05256 [quant-ph]} \BibitemShut
  {NoStop}%
%%CITATION = ARXIV:1607.05256;%%
\bibitem [{\citenamefont {Watrous}(2009)}]{johnw}%
  \BibitemOpen
  \bibfield  {author} {\bibinfo {author} {\bibfnamefont {J.}~\bibnamefont
  {Watrous}},\ }in\ \href@noop {} {\emph {\bibinfo {booktitle} {Encyclopedia of
  complexity and systems science}}}\ (\bibinfo  {publisher} {Springer},\
  \bibinfo {year} {2009})\ pp.\ \bibinfo {pages} {7174--7201}\BibitemShut
  {NoStop}%
\bibitem [{\citenamefont {Susskind}(2018{\natexlab{a}})}]{Susskind:2018pmk}%
  \BibitemOpen
  \bibfield  {author} {\bibinfo {author} {\bibfnamefont {L.}~\bibnamefont
  {Susskind}}\ }(\bibinfo {year} {2018})\ \Eprint
  {http://arxiv.org/abs/1810.11563} {arXiv:1810.11563 [hep-th]} \BibitemShut
  {NoStop}%
%%CITATION = ARXIV:1810.11563;%%
\bibitem [{\citenamefont {Susskind}(2016{\natexlab{a}})}]{Susskind:2014rva}%
  \BibitemOpen
  \bibfield  {author} {\bibinfo {author} {\bibfnamefont {L.}~\bibnamefont
  {Susskind}},\ }\href {\doibase 10.1002/prop.201500092} {\bibfield  {journal}
  {\bibinfo  {journal} {Fortsch. Phys.}\ }\textbf {\bibinfo {volume} {64}},\
  \bibinfo {pages} {24} (\bibinfo {year} {2016}{\natexlab{a}})},\ \Eprint
  {http://arxiv.org/abs/1403.5695} {arXiv:1403.5695 [hep-th]} \BibitemShut
  {NoStop}%
%%CITATION = ARXIV:1403.5695;%%
\bibitem [{\citenamefont {Stanford}\ and\ \citenamefont
  {Susskind}(2014)}]{Stanford:2014jda}%
  \BibitemOpen
  \bibfield  {author} {\bibinfo {author} {\bibfnamefont {D.}~\bibnamefont
  {Stanford}}\ and\ \bibinfo {author} {\bibfnamefont {L.}~\bibnamefont
  {Susskind}},\ }\href {\doibase 10.1103/PhysRevD.90.126007} {\bibfield
  {journal} {\bibinfo  {journal} {Phys. Rev.}\ }\textbf {\bibinfo {volume}
  {D90}},\ \bibinfo {pages} {126007} (\bibinfo {year} {2014})},\ \Eprint
  {http://arxiv.org/abs/1406.2678} {arXiv:1406.2678 [hep-th]} \BibitemShut
  {NoStop}%
%%CITATION = ARXIV:1406.2678;%%
\bibitem [{\citenamefont {Brown}\ \emph
  {et~al.}(2016{\natexlab{a}})\citenamefont {Brown}, \citenamefont {Roberts},
  \citenamefont {Susskind}, \citenamefont {Swingle},\ and\ \citenamefont
  {Zhao}}]{Brown:2015bva}%
  \BibitemOpen
  \bibfield  {author} {\bibinfo {author} {\bibfnamefont {A.~R.}\ \bibnamefont
  {Brown}}, \bibinfo {author} {\bibfnamefont {D.~A.}\ \bibnamefont {Roberts}},
  \bibinfo {author} {\bibfnamefont {L.}~\bibnamefont {Susskind}}, \bibinfo
  {author} {\bibfnamefont {B.}~\bibnamefont {Swingle}}, \ and\ \bibinfo
  {author} {\bibfnamefont {Y.}~\bibnamefont {Zhao}},\ }\href {\doibase
  10.1103/PhysRevLett.116.191301} {\bibfield  {journal} {\bibinfo  {journal}
  {Phys. Rev. Lett.}\ }\textbf {\bibinfo {volume} {116}},\ \bibinfo {pages}
  {191301} (\bibinfo {year} {2016}{\natexlab{a}})},\ \Eprint
  {http://arxiv.org/abs/1509.07876} {arXiv:1509.07876 [hep-th]} \BibitemShut
  {NoStop}%
%%CITATION = ARXIV:1509.07876;%%
\bibitem [{\citenamefont {Brown}\ \emph
  {et~al.}(2016{\natexlab{b}})\citenamefont {Brown}, \citenamefont {Roberts},
  \citenamefont {Susskind}, \citenamefont {Swingle},\ and\ \citenamefont
  {Zhao}}]{Brown:2015lvg}%
  \BibitemOpen
  \bibfield  {author} {\bibinfo {author} {\bibfnamefont {A.~R.}\ \bibnamefont
  {Brown}}, \bibinfo {author} {\bibfnamefont {D.~A.}\ \bibnamefont {Roberts}},
  \bibinfo {author} {\bibfnamefont {L.}~\bibnamefont {Susskind}}, \bibinfo
  {author} {\bibfnamefont {B.}~\bibnamefont {Swingle}}, \ and\ \bibinfo
  {author} {\bibfnamefont {Y.}~\bibnamefont {Zhao}},\ }\href {\doibase
  10.1103/PhysRevD.93.086006} {\bibfield  {journal} {\bibinfo  {journal} {Phys.
  Rev.}\ }\textbf {\bibinfo {volume} {D93}},\ \bibinfo {pages} {086006}
  (\bibinfo {year} {2016}{\natexlab{b}})},\ \Eprint
  {http://arxiv.org/abs/1512.04993} {arXiv:1512.04993 [hep-th]} \BibitemShut
  {NoStop}%
%%CITATION = ARXIV:1512.04993;%%
\bibitem [{\citenamefont {Couch}\ \emph {et~al.}(2017)\citenamefont {Couch},
  \citenamefont {Fischler},\ and\ \citenamefont {Nguyen}}]{Couch:2016exn}%
  \BibitemOpen
  \bibfield  {author} {\bibinfo {author} {\bibfnamefont {J.}~\bibnamefont
  {Couch}}, \bibinfo {author} {\bibfnamefont {W.}~\bibnamefont {Fischler}}, \
  and\ \bibinfo {author} {\bibfnamefont {P.~H.}\ \bibnamefont {Nguyen}},\
  }\href {\doibase 10.1007/JHEP03(2017)119} {\bibfield  {journal} {\bibinfo
  {journal} {JHEP}\ }\textbf {\bibinfo {volume} {03}},\ \bibinfo {pages} {119}
  (\bibinfo {year} {2017})},\ \Eprint {http://arxiv.org/abs/1610.02038}
  {arXiv:1610.02038 [hep-th]} \BibitemShut {NoStop}%
%%CITATION = ARXIV:1610.02038;%%
\bibitem [{\citenamefont {Susskind}(2016{\natexlab{b}})}]{Susskind:2014moa}%
  \BibitemOpen
  \bibfield  {author} {\bibinfo {author} {\bibfnamefont {L.}~\bibnamefont
  {Susskind}},\ }\href {\doibase 10.1002/prop.201500095} {\bibfield  {journal}
  {\bibinfo  {journal} {Fortsch. Phys.}\ }\textbf {\bibinfo {volume} {64}},\
  \bibinfo {pages} {49} (\bibinfo {year} {2016}{\natexlab{b}})},\ \Eprint
  {http://arxiv.org/abs/1411.0690} {arXiv:1411.0690 [hep-th]} \BibitemShut
  {NoStop}%
%%CITATION = ARXIV:1411.0690;%%
\bibitem [{\citenamefont {Susskind}\ and\ \citenamefont
  {Zhao}(2014)}]{Susskind:2014jwa}%
  \BibitemOpen
  \bibfield  {author} {\bibinfo {author} {\bibfnamefont {L.}~\bibnamefont
  {Susskind}}\ and\ \bibinfo {author} {\bibfnamefont {Y.}~\bibnamefont
  {Zhao}},\ }\href@noop {} {\  (\bibinfo {year} {2014})},\ \Eprint
  {http://arxiv.org/abs/1408.2823} {arXiv:1408.2823 [hep-th]} \BibitemShut
  {NoStop}%
%%CITATION = ARXIV:1408.2823;%%
\bibitem [{\citenamefont {Susskind}(2016{\natexlab{c}})}]{Susskind:2015toa}%
  \BibitemOpen
  \bibfield  {author} {\bibinfo {author} {\bibfnamefont {L.}~\bibnamefont
  {Susskind}},\ }\href {\doibase 10.1002/prop.201500091} {\bibfield  {journal}
  {\bibinfo  {journal} {Fortsch. Phys.}\ }\textbf {\bibinfo {volume} {64}},\
  \bibinfo {pages} {84} (\bibinfo {year} {2016}{\natexlab{c}})},\ \Eprint
  {http://arxiv.org/abs/1507.02287} {arXiv:1507.02287 [hep-th]} \BibitemShut
  {NoStop}%
%%CITATION = ARXIV:1507.02287;%%
\bibitem [{\citenamefont {Roberts}\ \emph {et~al.}(2015)\citenamefont
  {Roberts}, \citenamefont {Stanford},\ and\ \citenamefont
  {Susskind}}]{Roberts:2014isa}%
  \BibitemOpen
  \bibfield  {author} {\bibinfo {author} {\bibfnamefont {D.~A.}\ \bibnamefont
  {Roberts}}, \bibinfo {author} {\bibfnamefont {D.}~\bibnamefont {Stanford}}, \
  and\ \bibinfo {author} {\bibfnamefont {L.}~\bibnamefont {Susskind}},\ }\href
  {\doibase 10.1007/JHEP03(2015)051} {\bibfield  {journal} {\bibinfo  {journal}
  {JHEP}\ }\textbf {\bibinfo {volume} {03}},\ \bibinfo {pages} {051} (\bibinfo
  {year} {2015})},\ \Eprint {http://arxiv.org/abs/1409.8180} {arXiv:1409.8180
  [hep-th]} \BibitemShut {NoStop}%
%%CITATION = ARXIV:1409.8180;%%
\bibitem [{\citenamefont {Lehner}\ \emph {et~al.}(2016)\citenamefont {Lehner},
  \citenamefont {Myers}, \citenamefont {Poisson},\ and\ \citenamefont
  {Sorkin}}]{Lehner:2016vdi}%
  \BibitemOpen
  \bibfield  {author} {\bibinfo {author} {\bibfnamefont {L.}~\bibnamefont
  {Lehner}}, \bibinfo {author} {\bibfnamefont {R.~C.}\ \bibnamefont {Myers}},
  \bibinfo {author} {\bibfnamefont {E.}~\bibnamefont {Poisson}}, \ and\
  \bibinfo {author} {\bibfnamefont {R.~D.}\ \bibnamefont {Sorkin}},\ }\href
  {\doibase 10.1103/PhysRevD.94.084046} {\bibfield  {journal} {\bibinfo
  {journal} {Phys. Rev.}\ }\textbf {\bibinfo {volume} {D94}},\ \bibinfo {pages}
  {084046} (\bibinfo {year} {2016})},\ \Eprint
  {http://arxiv.org/abs/1609.00207} {arXiv:1609.00207 [hep-th]} \BibitemShut
  {NoStop}%
%%CITATION = ARXIV:1609.00207;%%
\bibitem [{\citenamefont {Cai}\ \emph {et~al.}(2016)\citenamefont {Cai},
  \citenamefont {Ruan}, \citenamefont {Wang}, \citenamefont {Yang},\ and\
  \citenamefont {Peng}}]{Cai:2016xho}%
  \BibitemOpen
  \bibfield  {author} {\bibinfo {author} {\bibfnamefont {R.-G.}\ \bibnamefont
  {Cai}}, \bibinfo {author} {\bibfnamefont {S.-M.}\ \bibnamefont {Ruan}},
  \bibinfo {author} {\bibfnamefont {S.-J.}\ \bibnamefont {Wang}}, \bibinfo
  {author} {\bibfnamefont {R.-Q.}\ \bibnamefont {Yang}}, \ and\ \bibinfo
  {author} {\bibfnamefont {R.-H.}\ \bibnamefont {Peng}},\ }\href {\doibase
  10.1007/JHEP09(2016)161} {\bibfield  {journal} {\bibinfo  {journal} {JHEP}\
  }\textbf {\bibinfo {volume} {09}},\ \bibinfo {pages} {161} (\bibinfo {year}
  {2016})},\ \Eprint {http://arxiv.org/abs/1606.08307} {arXiv:1606.08307
  [gr-qc]} \BibitemShut {NoStop}%
%%CITATION = ARXIV:1606.08307;%%
\bibitem [{\citenamefont {Reynolds}\ and\ \citenamefont
  {Ross}(2017{\natexlab{a}})}]{Reynolds:2016rvl}%
  \BibitemOpen
  \bibfield  {author} {\bibinfo {author} {\bibfnamefont {A.}~\bibnamefont
  {Reynolds}}\ and\ \bibinfo {author} {\bibfnamefont {S.~F.}\ \bibnamefont
  {Ross}},\ }\href {\doibase 10.1088/1361-6382/aa6925} {\bibfield  {journal}
  {\bibinfo  {journal} {Class. Quant. Grav.}\ }\textbf {\bibinfo {volume}
  {34}},\ \bibinfo {pages} {105004} (\bibinfo {year} {2017}{\natexlab{a}})},\
  \Eprint {http://arxiv.org/abs/1612.05439} {arXiv:1612.05439 [hep-th]}
  \BibitemShut {NoStop}%
%%CITATION = ARXIV:1612.05439;%%
\bibitem [{\citenamefont {Chapman}\ \emph {et~al.}(2017)\citenamefont
  {Chapman}, \citenamefont {Marrochio},\ and\ \citenamefont
  {Myers}}]{Chapman:2016hwi}%
  \BibitemOpen
  \bibfield  {author} {\bibinfo {author} {\bibfnamefont {S.}~\bibnamefont
  {Chapman}}, \bibinfo {author} {\bibfnamefont {H.}~\bibnamefont {Marrochio}},
  \ and\ \bibinfo {author} {\bibfnamefont {R.~C.}\ \bibnamefont {Myers}},\
  }\href {\doibase 10.1007/JHEP01(2017)062} {\bibfield  {journal} {\bibinfo
  {journal} {JHEP}\ }\textbf {\bibinfo {volume} {01}},\ \bibinfo {pages} {062}
  (\bibinfo {year} {2017})},\ \Eprint {http://arxiv.org/abs/1610.08063}
  {arXiv:1610.08063 [hep-th]} \BibitemShut {NoStop}%
%%CITATION = ARXIV:1610.08063;%%
\bibitem [{\citenamefont {Carmi}\ \emph
  {et~al.}(2017{\natexlab{a}})\citenamefont {Carmi}, \citenamefont {Myers},\
  and\ \citenamefont {Rath}}]{Carmi:2016wjl}%
  \BibitemOpen
  \bibfield  {author} {\bibinfo {author} {\bibfnamefont {D.}~\bibnamefont
  {Carmi}}, \bibinfo {author} {\bibfnamefont {R.~C.}\ \bibnamefont {Myers}}, \
  and\ \bibinfo {author} {\bibfnamefont {P.}~\bibnamefont {Rath}},\ }\href
  {\doibase 10.1007/JHEP03(2017)118} {\bibfield  {journal} {\bibinfo  {journal}
  {JHEP}\ }\textbf {\bibinfo {volume} {03}},\ \bibinfo {pages} {118} (\bibinfo
  {year} {2017}{\natexlab{a}})},\ \Eprint {http://arxiv.org/abs/1612.00433}
  {arXiv:1612.00433 [hep-th]} \BibitemShut {NoStop}%
%%CITATION = ARXIV:1612.00433;%%
\bibitem [{\citenamefont {Moosa}(2018)}]{Moosa:2017yvt}%
  \BibitemOpen
  \bibfield  {author} {\bibinfo {author} {\bibfnamefont {M.}~\bibnamefont
  {Moosa}},\ }\href {\doibase 10.1007/JHEP03(2018)031} {\bibfield  {journal}
  {\bibinfo  {journal} {JHEP}\ }\textbf {\bibinfo {volume} {03}},\ \bibinfo
  {pages} {031} (\bibinfo {year} {2018})},\ \Eprint
  {http://arxiv.org/abs/1711.02668} {arXiv:1711.02668 [hep-th]} \BibitemShut
  {NoStop}%
%%CITATION = ARXIV:1711.02668;%%
\bibitem [{\citenamefont {Couch}\ \emph {et~al.}(2018)\citenamefont {Couch},
  \citenamefont {Eccles}, \citenamefont {Fischler},\ and\ \citenamefont
  {Xiao}}]{Couch:2017yil}%
  \BibitemOpen
  \bibfield  {author} {\bibinfo {author} {\bibfnamefont {J.}~\bibnamefont
  {Couch}}, \bibinfo {author} {\bibfnamefont {S.}~\bibnamefont {Eccles}},
  \bibinfo {author} {\bibfnamefont {W.}~\bibnamefont {Fischler}}, \ and\
  \bibinfo {author} {\bibfnamefont {M.-L.}\ \bibnamefont {Xiao}},\ }\href
  {\doibase 10.1007/JHEP03(2018)108} {\bibfield  {journal} {\bibinfo  {journal}
  {JHEP}\ }\textbf {\bibinfo {volume} {03}},\ \bibinfo {pages} {108} (\bibinfo
  {year} {2018})},\ \Eprint {http://arxiv.org/abs/1710.07833} {arXiv:1710.07833
  [hep-th]} \BibitemShut {NoStop}%
%%CITATION = ARXIV:1710.07833;%%
\bibitem [{\citenamefont {Cai}\ \emph {et~al.}(2017)\citenamefont {Cai},
  \citenamefont {Sasaki},\ and\ \citenamefont {Wang}}]{Cai:2017sjv}%
  \BibitemOpen
  \bibfield  {author} {\bibinfo {author} {\bibfnamefont {R.-G.}\ \bibnamefont
  {Cai}}, \bibinfo {author} {\bibfnamefont {M.}~\bibnamefont {Sasaki}}, \ and\
  \bibinfo {author} {\bibfnamefont {S.-J.}\ \bibnamefont {Wang}},\ }\href
  {\doibase 10.1103/PhysRevD.95.124002} {\bibfield  {journal} {\bibinfo
  {journal} {Phys. Rev.}\ }\textbf {\bibinfo {volume} {D95}},\ \bibinfo {pages}
  {124002} (\bibinfo {year} {2017})},\ \Eprint
  {http://arxiv.org/abs/1702.06766} {arXiv:1702.06766 [gr-qc]} \BibitemShut
  {NoStop}%
%%CITATION = ARXIV:1702.06766;%%
\bibitem [{\citenamefont {Brown}\ and\ \citenamefont
  {Susskind}(2018)}]{Brown:2017jil}%
  \BibitemOpen
  \bibfield  {author} {\bibinfo {author} {\bibfnamefont {A.~R.}\ \bibnamefont
  {Brown}}\ and\ \bibinfo {author} {\bibfnamefont {L.}~\bibnamefont
  {Susskind}},\ }\href {\doibase 10.1103/PhysRevD.97.086015} {\bibfield
  {journal} {\bibinfo  {journal} {Phys. Rev.}\ }\textbf {\bibinfo {volume}
  {D97}},\ \bibinfo {pages} {086015} (\bibinfo {year} {2018})},\ \Eprint
  {http://arxiv.org/abs/1701.01107} {arXiv:1701.01107 [hep-th]} \BibitemShut
  {NoStop}%
%%CITATION = ARXIV:1701.01107;%%
\bibitem [{\citenamefont {Carmi}\ \emph
  {et~al.}(2017{\natexlab{b}})\citenamefont {Carmi}, \citenamefont {Chapman},
  \citenamefont {Marrochio}, \citenamefont {Myers},\ and\ \citenamefont
  {Sugishita}}]{Carmi:2017jqz}%
  \BibitemOpen
  \bibfield  {author} {\bibinfo {author} {\bibfnamefont {D.}~\bibnamefont
  {Carmi}}, \bibinfo {author} {\bibfnamefont {S.}~\bibnamefont {Chapman}},
  \bibinfo {author} {\bibfnamefont {H.}~\bibnamefont {Marrochio}}, \bibinfo
  {author} {\bibfnamefont {R.~C.}\ \bibnamefont {Myers}}, \ and\ \bibinfo
  {author} {\bibfnamefont {S.}~\bibnamefont {Sugishita}},\ }\href {\doibase
  10.1007/JHEP11(2017)188} {\bibfield  {journal} {\bibinfo  {journal} {JHEP}\
  }\textbf {\bibinfo {volume} {11}},\ \bibinfo {pages} {188} (\bibinfo {year}
  {2017}{\natexlab{b}})},\ \Eprint {http://arxiv.org/abs/1709.10184}
  {arXiv:1709.10184 [hep-th]} \BibitemShut {NoStop}%
%%CITATION = ARXIV:1709.10184;%%
\bibitem [{\citenamefont {Swingle}\ and\ \citenamefont
  {Wang}(2018)}]{Swingle:2017zcd}%
  \BibitemOpen
  \bibfield  {author} {\bibinfo {author} {\bibfnamefont {B.}~\bibnamefont
  {Swingle}}\ and\ \bibinfo {author} {\bibfnamefont {Y.}~\bibnamefont {Wang}},\
  }\href {\doibase 10.1007/JHEP09(2018)106} {\bibfield  {journal} {\bibinfo
  {journal} {JHEP}\ }\textbf {\bibinfo {volume} {09}},\ \bibinfo {pages} {106}
  (\bibinfo {year} {2018})},\ \Eprint {http://arxiv.org/abs/1712.09826}
  {arXiv:1712.09826 [hep-th]} \BibitemShut {NoStop}%
%%CITATION = ARXIV:1712.09826;%%
\bibitem [{\citenamefont {Flory}(2017)}]{Flory:2017ftd}%
  \BibitemOpen
  \bibfield  {author} {\bibinfo {author} {\bibfnamefont {M.}~\bibnamefont
  {Flory}},\ }\href {\doibase 10.1007/JHEP06(2017)131} {\bibfield  {journal}
  {\bibinfo  {journal} {JHEP}\ }\textbf {\bibinfo {volume} {06}},\ \bibinfo
  {pages} {131} (\bibinfo {year} {2017})},\ \Eprint
  {http://arxiv.org/abs/1702.06386} {arXiv:1702.06386 [hep-th]} \BibitemShut
  {NoStop}%
%%CITATION = ARXIV:1702.06386;%%
\bibitem [{\citenamefont {Zhao}(2018)}]{Zhao:2017isy}%
  \BibitemOpen
  \bibfield  {author} {\bibinfo {author} {\bibfnamefont {Y.}~\bibnamefont
  {Zhao}},\ }\href {\doibase 10.1103/PhysRevD.97.126007} {\bibfield  {journal}
  {\bibinfo  {journal} {Phys. Rev.}\ }\textbf {\bibinfo {volume} {D97}},\
  \bibinfo {pages} {126007} (\bibinfo {year} {2018})},\ \Eprint
  {http://arxiv.org/abs/1711.03125} {arXiv:1711.03125 [hep-th]} \BibitemShut
  {NoStop}%
%%CITATION = ARXIV:1711.03125;%%
\bibitem [{\citenamefont {Abt}\ \emph {et~al.}(2018)\citenamefont {Abt},
  \citenamefont {Erdmenger}, \citenamefont {Hinrichsen}, \citenamefont
  {Melby-Thompson}, \citenamefont {Meyer}, \citenamefont {Northe},\ and\
  \citenamefont {Reyes}}]{Abt:2017pmf}%
  \BibitemOpen
  \bibfield  {author} {\bibinfo {author} {\bibfnamefont {R.}~\bibnamefont
  {Abt}}, \bibinfo {author} {\bibfnamefont {J.}~\bibnamefont {Erdmenger}},
  \bibinfo {author} {\bibfnamefont {H.}~\bibnamefont {Hinrichsen}}, \bibinfo
  {author} {\bibfnamefont {C.~M.}\ \bibnamefont {Melby-Thompson}}, \bibinfo
  {author} {\bibfnamefont {R.}~\bibnamefont {Meyer}}, \bibinfo {author}
  {\bibfnamefont {C.}~\bibnamefont {Northe}}, \ and\ \bibinfo {author}
  {\bibfnamefont {I.~A.}\ \bibnamefont {Reyes}},\ }\href {\doibase
  10.1002/prop.201800034} {\bibfield  {journal} {\bibinfo  {journal} {Fortsch.
  Phys.}\ }\textbf {\bibinfo {volume} {66}},\ \bibinfo {pages} {1800034}
  (\bibinfo {year} {2018})},\ \Eprint {http://arxiv.org/abs/1710.01327}
  {arXiv:1710.01327 [hep-th]} \BibitemShut {NoStop}%
%%CITATION = ARXIV:1710.01327;%%
\bibitem [{\citenamefont {Abt}\ \emph {et~al.}(2019)\citenamefont {Abt},
  \citenamefont {Erdmenger}, \citenamefont {Gerbershagen}, \citenamefont
  {Melby-Thompson},\ and\ \citenamefont {Northe}}]{Abt:2018ywl}%
  \BibitemOpen
  \bibfield  {author} {\bibinfo {author} {\bibfnamefont {R.}~\bibnamefont
  {Abt}}, \bibinfo {author} {\bibfnamefont {J.}~\bibnamefont {Erdmenger}},
  \bibinfo {author} {\bibfnamefont {M.}~\bibnamefont {Gerbershagen}}, \bibinfo
  {author} {\bibfnamefont {C.~M.}\ \bibnamefont {Melby-Thompson}}, \ and\
  \bibinfo {author} {\bibfnamefont {C.}~\bibnamefont {Northe}},\ }\href
  {\doibase 10.1007/JHEP01(2019)012} {\bibfield  {journal} {\bibinfo  {journal}
  {JHEP}\ }\textbf {\bibinfo {volume} {01}},\ \bibinfo {pages} {012} (\bibinfo
  {year} {2019})},\ \Eprint {http://arxiv.org/abs/1805.10298} {arXiv:1805.10298
  [hep-th]} \BibitemShut {NoStop}%
%%CITATION = ARXIV:1805.10298;%%
\bibitem [{\citenamefont {Alishahiha}\ \emph {et~al.}(2018)\citenamefont
  {Alishahiha}, \citenamefont {Faraji~Astaneh}, \citenamefont
  {Mohammadi~Mozaffar},\ and\ \citenamefont {Mollabashi}}]{Alishahiha:2018tep}%
  \BibitemOpen
  \bibfield  {author} {\bibinfo {author} {\bibfnamefont {M.}~\bibnamefont
  {Alishahiha}}, \bibinfo {author} {\bibfnamefont {A.}~\bibnamefont
  {Faraji~Astaneh}}, \bibinfo {author} {\bibfnamefont {M.~R.}\ \bibnamefont
  {Mohammadi~Mozaffar}}, \ and\ \bibinfo {author} {\bibfnamefont
  {A.}~\bibnamefont {Mollabashi}},\ }\href {\doibase 10.1007/JHEP07(2018)042}
  {\bibfield  {journal} {\bibinfo  {journal} {JHEP}\ }\textbf {\bibinfo
  {volume} {07}},\ \bibinfo {pages} {042} (\bibinfo {year} {2018})},\ \Eprint
  {http://arxiv.org/abs/1802.06740} {arXiv:1802.06740 [hep-th]} \BibitemShut
  {NoStop}%
%%CITATION = ARXIV:1802.06740;%%
\bibitem [{\citenamefont {An}\ and\ \citenamefont {Peng}(2018)}]{An:2018xhv}%
  \BibitemOpen
  \bibfield  {author} {\bibinfo {author} {\bibfnamefont {Y.-S.}\ \bibnamefont
  {An}}\ and\ \bibinfo {author} {\bibfnamefont {R.-H.}\ \bibnamefont {Peng}},\
  }\href {\doibase 10.1103/PhysRevD.97.066022} {\bibfield  {journal} {\bibinfo
  {journal} {Phys. Rev.}\ }\textbf {\bibinfo {volume} {D97}},\ \bibinfo {pages}
  {066022} (\bibinfo {year} {2018})},\ \Eprint
  {http://arxiv.org/abs/1801.03638} {arXiv:1801.03638 [hep-th]} \BibitemShut
  {NoStop}%
%%CITATION = ARXIV:1801.03638;%%
\bibitem [{\citenamefont {Fu}\ \emph {et~al.}(2018)\citenamefont {Fu},
  \citenamefont {Maloney}, \citenamefont {Marolf}, \citenamefont {Maxfield},\
  and\ \citenamefont {Wang}}]{Fu:2018kcp}%
  \BibitemOpen
  \bibfield  {author} {\bibinfo {author} {\bibfnamefont {Z.}~\bibnamefont
  {Fu}}, \bibinfo {author} {\bibfnamefont {A.}~\bibnamefont {Maloney}},
  \bibinfo {author} {\bibfnamefont {D.}~\bibnamefont {Marolf}}, \bibinfo
  {author} {\bibfnamefont {H.}~\bibnamefont {Maxfield}}, \ and\ \bibinfo
  {author} {\bibfnamefont {Z.}~\bibnamefont {Wang}},\ }\href {\doibase
  10.1007/JHEP02(2018)072} {\bibfield  {journal} {\bibinfo  {journal} {JHEP}\
  }\textbf {\bibinfo {volume} {02}},\ \bibinfo {pages} {072} (\bibinfo {year}
  {2018})},\ \Eprint {http://arxiv.org/abs/1801.01137} {arXiv:1801.01137
  [hep-th]} \BibitemShut {NoStop}%
%%CITATION = ARXIV:1801.01137;%%
\bibitem [{\citenamefont {Mahapatra}\ and\ \citenamefont
  {Roy}(2018)}]{Mahapatra:2018gig}%
  \BibitemOpen
  \bibfield  {author} {\bibinfo {author} {\bibfnamefont {S.}~\bibnamefont
  {Mahapatra}}\ and\ \bibinfo {author} {\bibfnamefont {P.}~\bibnamefont
  {Roy}},\ }\href {\doibase 10.1007/JHEP11(2018)138} {\bibfield  {journal}
  {\bibinfo  {journal} {JHEP}\ }\textbf {\bibinfo {volume} {11}},\ \bibinfo
  {pages} {138} (\bibinfo {year} {2018})},\ \Eprint
  {http://arxiv.org/abs/1808.09917} {arXiv:1808.09917 [hep-th]} \BibitemShut
  {NoStop}%
%%CITATION = ARXIV:1808.09917;%%
\bibitem [{\citenamefont {Chapman}\ \emph
  {et~al.}(2018{\natexlab{a}})\citenamefont {Chapman}, \citenamefont
  {Marrochio},\ and\ \citenamefont {Myers}}]{Chapman:2018dem}%
  \BibitemOpen
  \bibfield  {author} {\bibinfo {author} {\bibfnamefont {S.}~\bibnamefont
  {Chapman}}, \bibinfo {author} {\bibfnamefont {H.}~\bibnamefont {Marrochio}},
  \ and\ \bibinfo {author} {\bibfnamefont {R.~C.}\ \bibnamefont {Myers}},\
  }\href {\doibase 10.1007/JHEP06(2018)046} {\bibfield  {journal} {\bibinfo
  {journal} {JHEP}\ }\textbf {\bibinfo {volume} {06}},\ \bibinfo {pages} {046}
  (\bibinfo {year} {2018}{\natexlab{a}})},\ \Eprint
  {http://arxiv.org/abs/1804.07410} {arXiv:1804.07410 [hep-th]} \BibitemShut
  {NoStop}%
%%CITATION = ARXIV:1804.07410;%%
\bibitem [{\citenamefont {Chapman}\ \emph
  {et~al.}(2018{\natexlab{b}})\citenamefont {Chapman}, \citenamefont
  {Marrochio},\ and\ \citenamefont {Myers}}]{Chapman:2018lsv}%
  \BibitemOpen
  \bibfield  {author} {\bibinfo {author} {\bibfnamefont {S.}~\bibnamefont
  {Chapman}}, \bibinfo {author} {\bibfnamefont {H.}~\bibnamefont {Marrochio}},
  \ and\ \bibinfo {author} {\bibfnamefont {R.~C.}\ \bibnamefont {Myers}},\
  }\href {\doibase 10.1007/JHEP06(2018)114} {\bibfield  {journal} {\bibinfo
  {journal} {JHEP}\ }\textbf {\bibinfo {volume} {06}},\ \bibinfo {pages} {114}
  (\bibinfo {year} {2018}{\natexlab{b}})},\ \Eprint
  {http://arxiv.org/abs/1805.07262} {arXiv:1805.07262 [hep-th]} \BibitemShut
  {NoStop}%
%%CITATION = ARXIV:1805.07262;%%
\bibitem [{\citenamefont {Cano}\ \emph {et~al.}(2018)\citenamefont {Cano},
  \citenamefont {Hennigar},\ and\ \citenamefont {Marrochio}}]{Cano:2018aqi}%
  \BibitemOpen
  \bibfield  {author} {\bibinfo {author} {\bibfnamefont {P.~A.}\ \bibnamefont
  {Cano}}, \bibinfo {author} {\bibfnamefont {R.~A.}\ \bibnamefont {Hennigar}},
  \ and\ \bibinfo {author} {\bibfnamefont {H.}~\bibnamefont {Marrochio}},\
  }\href {\doibase 10.1103/PhysRevLett.121.121602} {\bibfield  {journal}
  {\bibinfo  {journal} {Phys. Rev. Lett.}\ }\textbf {\bibinfo {volume} {121}},\
  \bibinfo {pages} {121602} (\bibinfo {year} {2018})},\ \Eprint
  {http://arxiv.org/abs/1803.02795} {arXiv:1803.02795 [hep-th]} \BibitemShut
  {NoStop}%
%%CITATION = ARXIV:1803.02795;%%
\bibitem [{\citenamefont {Barbon}\ and\ \citenamefont
  {Martin-Garcia}(2018)}]{Barbon:2018mxk}%
  \BibitemOpen
  \bibfield  {author} {\bibinfo {author} {\bibfnamefont {J.~L.~F.}\
  \bibnamefont {Barbon}}\ and\ \bibinfo {author} {\bibfnamefont
  {J.}~\bibnamefont {Martin-Garcia}},\ }\href {\doibase
  10.1007/JHEP06(2018)132} {\bibfield  {journal} {\bibinfo  {journal} {JHEP}\
  }\textbf {\bibinfo {volume} {06}},\ \bibinfo {pages} {132} (\bibinfo {year}
  {2018})},\ \Eprint {http://arxiv.org/abs/1805.05291} {arXiv:1805.05291
  [hep-th]} \BibitemShut {NoStop}%
%%CITATION = ARXIV:1805.05291;%%
\bibitem [{\citenamefont {Susskind}(2018{\natexlab{b}})}]{Susskind:2018fmx}%
  \BibitemOpen
  \bibfield  {author} {\bibinfo {author} {\bibfnamefont {L.}~\bibnamefont
  {Susskind}},\ }\href@noop {} {\  (\bibinfo {year} {2018}{\natexlab{b}})},\
  \Eprint {http://arxiv.org/abs/1802.02175} {arXiv:1802.02175 [hep-th]}
  \BibitemShut {NoStop}%
%%CITATION = ARXIV:1802.02175;%%
\bibitem [{\citenamefont {Susskind}(2018{\natexlab{c}})}]{Susskind:2018tei}%
  \BibitemOpen
  \bibfield  {author} {\bibinfo {author} {\bibfnamefont {L.}~\bibnamefont
  {Susskind}},\ }\href@noop {} {\  (\bibinfo {year} {2018}{\natexlab{c}})},\
  \Eprint {http://arxiv.org/abs/1802.01198} {arXiv:1802.01198 [hep-th]}
  \BibitemShut {NoStop}%
%%CITATION = ARXIV:1802.01198;%%
\bibitem [{\citenamefont {Cooper}\ \emph {et~al.}(2018)\citenamefont {Cooper},
  \citenamefont {Rozali}, \citenamefont {Swingle}, \citenamefont
  {Van~Raamsdonk}, \citenamefont {Waddell},\ and\ \citenamefont
  {Wakeham}}]{Cooper:2018cmb}%
  \BibitemOpen
  \bibfield  {author} {\bibinfo {author} {\bibfnamefont {S.}~\bibnamefont
  {Cooper}}, \bibinfo {author} {\bibfnamefont {M.}~\bibnamefont {Rozali}},
  \bibinfo {author} {\bibfnamefont {B.}~\bibnamefont {Swingle}}, \bibinfo
  {author} {\bibfnamefont {M.}~\bibnamefont {Van~Raamsdonk}}, \bibinfo {author}
  {\bibfnamefont {C.}~\bibnamefont {Waddell}}, \ and\ \bibinfo {author}
  {\bibfnamefont {D.}~\bibnamefont {Wakeham}},\ }\href@noop {} {\  (\bibinfo
  {year} {2018})},\ \Eprint {http://arxiv.org/abs/1810.10601} {arXiv:1810.10601
  [hep-th]} \BibitemShut {NoStop}%
%%CITATION = ARXIV:1810.10601;%%
\bibitem [{\citenamefont {Numasawa}(2018)}]{Numasawa:2018grg}%
  \BibitemOpen
  \bibfield  {author} {\bibinfo {author} {\bibfnamefont {T.}~\bibnamefont
  {Numasawa}},\ }\href@noop {} {\  (\bibinfo {year} {2018})},\ \Eprint
  {http://arxiv.org/abs/1811.03597} {arXiv:1811.03597 [hep-th]} \BibitemShut
  {NoStop}%
%%CITATION = ARXIV:1811.03597;%%
\bibitem [{\citenamefont {Brown}\ \emph {et~al.}(2018)\citenamefont {Brown},
  \citenamefont {Gharibyan}, \citenamefont {Streicher}, \citenamefont
  {Susskind}, \citenamefont {Thorlacius},\ and\ \citenamefont
  {Zhao}}]{Brown:2018kvn}%
  \BibitemOpen
  \bibfield  {author} {\bibinfo {author} {\bibfnamefont {A.~R.}\ \bibnamefont
  {Brown}}, \bibinfo {author} {\bibfnamefont {H.}~\bibnamefont {Gharibyan}},
  \bibinfo {author} {\bibfnamefont {A.}~\bibnamefont {Streicher}}, \bibinfo
  {author} {\bibfnamefont {L.}~\bibnamefont {Susskind}}, \bibinfo {author}
  {\bibfnamefont {L.}~\bibnamefont {Thorlacius}}, \ and\ \bibinfo {author}
  {\bibfnamefont {Y.}~\bibnamefont {Zhao}},\ }\href {\doibase
  10.1103/PhysRevD.98.126016} {\bibfield  {journal} {\bibinfo  {journal} {Phys.
  Rev.}\ }\textbf {\bibinfo {volume} {D98}},\ \bibinfo {pages} {126016}
  (\bibinfo {year} {2018})},\ \Eprint {http://arxiv.org/abs/1804.04156}
  {arXiv:1804.04156 [hep-th]} \BibitemShut {NoStop}%
%%CITATION = ARXIV:1804.04156;%%
\bibitem [{\citenamefont {Goto}\ \emph {et~al.}(2019)\citenamefont {Goto},
  \citenamefont {Marrochio}, \citenamefont {Myers}, \citenamefont {Queimada},\
  and\ \citenamefont {Yoshida}}]{Goto:2018iay}%
  \BibitemOpen
  \bibfield  {author} {\bibinfo {author} {\bibfnamefont {K.}~\bibnamefont
  {Goto}}, \bibinfo {author} {\bibfnamefont {H.}~\bibnamefont {Marrochio}},
  \bibinfo {author} {\bibfnamefont {R.~C.}\ \bibnamefont {Myers}}, \bibinfo
  {author} {\bibfnamefont {L.}~\bibnamefont {Queimada}}, \ and\ \bibinfo
  {author} {\bibfnamefont {B.}~\bibnamefont {Yoshida}},\ }\href {\doibase
  10.1007/JHEP02(2019)160} {\bibfield  {journal} {\bibinfo  {journal} {JHEP}\
  }\textbf {\bibinfo {volume} {02}},\ \bibinfo {pages} {160} (\bibinfo {year}
  {2019})},\ \Eprint {http://arxiv.org/abs/1901.00014} {arXiv:1901.00014
  [hep-th]} \BibitemShut {NoStop}%
%%CITATION = ARXIV:1901.00014;%%
\bibitem [{\citenamefont {Ag\'{o}n}\ \emph {et~al.}(2018)\citenamefont
  {Ag\'{o}n}, \citenamefont {Headrick},\ and\ \citenamefont
  {Swingle}}]{Agon:2018zso}%
  \BibitemOpen
  \bibfield  {author} {\bibinfo {author} {\bibfnamefont {C.~A.}\ \bibnamefont
  {Ag\'{o}n}}, \bibinfo {author} {\bibfnamefont {M.}~\bibnamefont {Headrick}},
  \ and\ \bibinfo {author} {\bibfnamefont {B.}~\bibnamefont {Swingle}},\
  }\href@noop {} {\  (\bibinfo {year} {2018})},\ \Eprint
  {http://arxiv.org/abs/1804.01561} {arXiv:1804.01561 [hep-th]} \BibitemShut
  {NoStop}%
%%CITATION = ARXIV:1804.01561;%%
\bibitem [{\citenamefont {Chapman}\ \emph
  {et~al.}(2018{\natexlab{c}})\citenamefont {Chapman}, \citenamefont {Ge},\
  and\ \citenamefont {Policastro}}]{Chapman:2018bqj}%
  \BibitemOpen
  \bibfield  {author} {\bibinfo {author} {\bibfnamefont {S.}~\bibnamefont
  {Chapman}}, \bibinfo {author} {\bibfnamefont {D.}~\bibnamefont {Ge}}, \ and\
  \bibinfo {author} {\bibfnamefont {G.}~\bibnamefont {Policastro}},\
  }\href@noop {} {\  (\bibinfo {year} {2018}{\natexlab{c}})},\ \Eprint
  {http://arxiv.org/abs/1811.12549} {arXiv:1811.12549 [hep-th]} \BibitemShut
  {NoStop}%
%%CITATION = ARXIV:1811.12549;%%
\bibitem [{\citenamefont {Flory}\ and\ \citenamefont
  {Miekley}(2018)}]{Flory:2018akz}%
  \BibitemOpen
  \bibfield  {author} {\bibinfo {author} {\bibfnamefont {M.}~\bibnamefont
  {Flory}}\ and\ \bibinfo {author} {\bibfnamefont {N.}~\bibnamefont
  {Miekley}},\ }\href@noop {} {\  (\bibinfo {year} {2018})},\ \Eprint
  {http://arxiv.org/abs/1806.08376} {arXiv:1806.08376 [hep-th]} \BibitemShut
  {NoStop}%
%%CITATION = ARXIV:1806.08376;%%
\bibitem [{\citenamefont {Flory}(2019)}]{Flory:2019kah}%
  \BibitemOpen
  \bibfield  {author} {\bibinfo {author} {\bibfnamefont {M.}~\bibnamefont
  {Flory}},\ }\href@noop {} {\  (\bibinfo {year} {2019})},\ \Eprint
  {http://arxiv.org/abs/1902.06499} {arXiv:1902.06499 [hep-th]} \BibitemShut
  {NoStop}%
%%CITATION = ARXIV:1902.06499;%%
\bibitem [{\citenamefont {Belin}\ \emph {et~al.}(2018)\citenamefont {Belin},
  \citenamefont {Lewkowycz},\ and\ \citenamefont {S\'arosi}}]{Belin:2018bpg}%
  \BibitemOpen
  \bibfield  {author} {\bibinfo {author} {\bibfnamefont {A.}~\bibnamefont
  {Belin}}, \bibinfo {author} {\bibfnamefont {A.}~\bibnamefont {Lewkowycz}}, \
  and\ \bibinfo {author} {\bibfnamefont {G.}~\bibnamefont {S\'arosi}},\
  }\href@noop {} {\  (\bibinfo {year} {2018})},\ \Eprint
  {http://arxiv.org/abs/1811.03097} {arXiv:1811.03097 [hep-th]} \BibitemShut
  {NoStop}%
%%CITATION = ARXIV:1811.03097;%%
\bibitem [{\citenamefont {Belin}\ \emph {et~al.}(2019)\citenamefont {Belin},
  \citenamefont {Lewkowycz},\ and\ \citenamefont {S\'arosi}}]{Belin:2018fxe}%
  \BibitemOpen
  \bibfield  {author} {\bibinfo {author} {\bibfnamefont {A.}~\bibnamefont
  {Belin}}, \bibinfo {author} {\bibfnamefont {A.}~\bibnamefont {Lewkowycz}}, \
  and\ \bibinfo {author} {\bibfnamefont {G.}~\bibnamefont {S\'arosi}},\ }\href
  {\doibase 10.1016/j.physletb.2018.10.071} {\bibfield  {journal} {\bibinfo
  {journal} {Phys. Lett.}\ }\textbf {\bibinfo {volume} {B789}},\ \bibinfo
  {pages} {71} (\bibinfo {year} {2019})},\ \Eprint
  {http://arxiv.org/abs/1806.10144} {arXiv:1806.10144 [hep-th]} \BibitemShut
  {NoStop}%
%%CITATION = ARXIV:1806.10144;%%
\bibitem [{\citenamefont {Bhattacharyya}\ \emph
  {et~al.}(2018{\natexlab{a}})\citenamefont {Bhattacharyya}, \citenamefont
  {Caputa}, \citenamefont {Das}, \citenamefont {Kundu}, \citenamefont
  {Miyaji},\ and\ \citenamefont {Takayanagi}}]{Bhattacharyya:2018wym}%
  \BibitemOpen
  \bibfield  {author} {\bibinfo {author} {\bibfnamefont {A.}~\bibnamefont
  {Bhattacharyya}}, \bibinfo {author} {\bibfnamefont {P.}~\bibnamefont
  {Caputa}}, \bibinfo {author} {\bibfnamefont {S.~R.}\ \bibnamefont {Das}},
  \bibinfo {author} {\bibfnamefont {N.}~\bibnamefont {Kundu}}, \bibinfo
  {author} {\bibfnamefont {M.}~\bibnamefont {Miyaji}}, \ and\ \bibinfo {author}
  {\bibfnamefont {T.}~\bibnamefont {Takayanagi}},\ }\href {\doibase
  10.1007/JHEP07(2018)086} {\bibfield  {journal} {\bibinfo  {journal} {JHEP}\
  }\textbf {\bibinfo {volume} {07}},\ \bibinfo {pages} {086} (\bibinfo {year}
  {2018}{\natexlab{a}})},\ \Eprint {http://arxiv.org/abs/1804.01999}
  {arXiv:1804.01999 [hep-th]} \BibitemShut {NoStop}%
%%CITATION = ARXIV:1804.01999;%%
\bibitem [{\citenamefont {Takayanagi}(2018)}]{Takayanagi:2018pml}%
  \BibitemOpen
  \bibfield  {author} {\bibinfo {author} {\bibfnamefont {T.}~\bibnamefont
  {Takayanagi}},\ }\href {\doibase 10.1007/JHEP12(2018)048} {\bibfield
  {journal} {\bibinfo  {journal} {JHEP}\ }\textbf {\bibinfo {volume} {12}},\
  \bibinfo {pages} {048} (\bibinfo {year} {2018})},\ \Eprint
  {http://arxiv.org/abs/1808.09072} {arXiv:1808.09072 [hep-th]} \BibitemShut
  {NoStop}%
%%CITATION = ARXIV:1808.09072;%%
\bibitem [{\citenamefont {Caputa}\ \emph
  {et~al.}(2017{\natexlab{a}})\citenamefont {Caputa}, \citenamefont {Kundu},
  \citenamefont {Miyaji}, \citenamefont {Takayanagi},\ and\ \citenamefont
  {Watanabe}}]{Caputa:2017urj}%
  \BibitemOpen
  \bibfield  {author} {\bibinfo {author} {\bibfnamefont {P.}~\bibnamefont
  {Caputa}}, \bibinfo {author} {\bibfnamefont {N.}~\bibnamefont {Kundu}},
  \bibinfo {author} {\bibfnamefont {M.}~\bibnamefont {Miyaji}}, \bibinfo
  {author} {\bibfnamefont {T.}~\bibnamefont {Takayanagi}}, \ and\ \bibinfo
  {author} {\bibfnamefont {K.}~\bibnamefont {Watanabe}},\ }\href {\doibase
  10.1103/PhysRevLett.119.071602} {\bibfield  {journal} {\bibinfo  {journal}
  {Phys. Rev. Lett.}\ }\textbf {\bibinfo {volume} {119}},\ \bibinfo {pages}
  {071602} (\bibinfo {year} {2017}{\natexlab{a}})},\ \Eprint
  {http://arxiv.org/abs/1703.00456} {arXiv:1703.00456 [hep-th]} \BibitemShut
  {NoStop}%
%%CITATION = ARXIV:1703.00456;%%
\bibitem [{\citenamefont {Caputa}\ \emph
  {et~al.}(2017{\natexlab{b}})\citenamefont {Caputa}, \citenamefont {Kundu},
  \citenamefont {Miyaji}, \citenamefont {Takayanagi},\ and\ \citenamefont
  {Watanabe}}]{Caputa:2017yrh}%
  \BibitemOpen
  \bibfield  {author} {\bibinfo {author} {\bibfnamefont {P.}~\bibnamefont
  {Caputa}}, \bibinfo {author} {\bibfnamefont {N.}~\bibnamefont {Kundu}},
  \bibinfo {author} {\bibfnamefont {M.}~\bibnamefont {Miyaji}}, \bibinfo
  {author} {\bibfnamefont {T.}~\bibnamefont {Takayanagi}}, \ and\ \bibinfo
  {author} {\bibfnamefont {K.}~\bibnamefont {Watanabe}},\ }\href {\doibase
  10.1007/JHEP11(2017)097} {\bibfield  {journal} {\bibinfo  {journal} {JHEP}\
  }\textbf {\bibinfo {volume} {11}},\ \bibinfo {pages} {097} (\bibinfo {year}
  {2017}{\natexlab{b}})},\ \Eprint {http://arxiv.org/abs/1706.07056}
  {arXiv:1706.07056 [hep-th]} \BibitemShut {NoStop}%
%%CITATION = ARXIV:1706.07056;%%
\bibitem [{\citenamefont {Jefferson}\ and\ \citenamefont {Myers}(2017)}]{Jeff}%
  \BibitemOpen
  \bibfield  {author} {\bibinfo {author} {\bibfnamefont {R.~A.}\ \bibnamefont
  {Jefferson}}\ and\ \bibinfo {author} {\bibfnamefont {R.~C.}\ \bibnamefont
  {Myers}},\ }\href {\doibase 10.1007/JHEP10(2017)107} {\bibfield  {journal}
  {\bibinfo  {journal} {JHEP}\ }\textbf {\bibinfo {volume} {10}},\ \bibinfo
  {pages} {107} (\bibinfo {year} {2017})},\ \Eprint
  {http://arxiv.org/abs/1707.08570} {arXiv:1707.08570 [hep-th]} \BibitemShut
  {NoStop}%
%%CITATION = ARXIV:1707.08570;%%
\bibitem [{\citenamefont {Chapman}\ \emph
  {et~al.}(2018{\natexlab{d}})\citenamefont {Chapman}, \citenamefont {Heller},
  \citenamefont {Marrochio},\ and\ \citenamefont
  {Pastawski}}]{Chapman:2017rqy}%
  \BibitemOpen
  \bibfield  {author} {\bibinfo {author} {\bibfnamefont {S.}~\bibnamefont
  {Chapman}}, \bibinfo {author} {\bibfnamefont {M.~P.}\ \bibnamefont {Heller}},
  \bibinfo {author} {\bibfnamefont {H.}~\bibnamefont {Marrochio}}, \ and\
  \bibinfo {author} {\bibfnamefont {F.}~\bibnamefont {Pastawski}},\ }\href
  {\doibase 10.1103/PhysRevLett.120.121602} {\bibfield  {journal} {\bibinfo
  {journal} {Phys. Rev. Lett.}\ }\textbf {\bibinfo {volume} {120}},\ \bibinfo
  {pages} {121602} (\bibinfo {year} {2018}{\natexlab{d}})},\ \Eprint
  {http://arxiv.org/abs/1707.08582} {arXiv:1707.08582 [hep-th]} \BibitemShut
  {NoStop}%
%%CITATION = ARXIV:1707.08582;%%
\bibitem [{\citenamefont {Hackl}\ and\ \citenamefont
  {Myers}(2018)}]{Hackl:2018ptj}%
  \BibitemOpen
  \bibfield  {author} {\bibinfo {author} {\bibfnamefont {L.}~\bibnamefont
  {Hackl}}\ and\ \bibinfo {author} {\bibfnamefont {R.~C.}\ \bibnamefont
  {Myers}},\ }\href {\doibase 10.1007/JHEP07(2018)139} {\bibfield  {journal}
  {\bibinfo  {journal} {JHEP}\ }\textbf {\bibinfo {volume} {07}},\ \bibinfo
  {pages} {139} (\bibinfo {year} {2018})},\ \Eprint
  {http://arxiv.org/abs/1803.10638} {arXiv:1803.10638 [hep-th]} \BibitemShut
  {NoStop}%
%%CITATION = ARXIV:1803.10638;%%
\bibitem [{\citenamefont {Khan}\ \emph {et~al.}(2018)\citenamefont {Khan},
  \citenamefont {Krishnan},\ and\ \citenamefont {Sharma}}]{Khan:2018rzm}%
  \BibitemOpen
  \bibfield  {author} {\bibinfo {author} {\bibfnamefont {R.}~\bibnamefont
  {Khan}}, \bibinfo {author} {\bibfnamefont {C.}~\bibnamefont {Krishnan}}, \
  and\ \bibinfo {author} {\bibfnamefont {S.}~\bibnamefont {Sharma}},\ }\href
  {\doibase 10.1103/PhysRevD.98.126001} {\bibfield  {journal} {\bibinfo
  {journal} {Phys. Rev.}\ }\textbf {\bibinfo {volume} {D98}},\ \bibinfo {pages}
  {126001} (\bibinfo {year} {2018})},\ \Eprint
  {http://arxiv.org/abs/1801.07620} {arXiv:1801.07620 [hep-th]} \BibitemShut
  {NoStop}%
%%CITATION = ARXIV:1801.07620;%%
\bibitem [{\citenamefont {Guo}\ \emph {et~al.}(2018)\citenamefont {Guo},
  \citenamefont {Hernandez}, \citenamefont {Myers},\ and\ \citenamefont
  {Ruan}}]{cohere}%
  \BibitemOpen
  \bibfield  {author} {\bibinfo {author} {\bibfnamefont {M.}~\bibnamefont
  {Guo}}, \bibinfo {author} {\bibfnamefont {J.}~\bibnamefont {Hernandez}},
  \bibinfo {author} {\bibfnamefont {R.~C.}\ \bibnamefont {Myers}}, \ and\
  \bibinfo {author} {\bibfnamefont {S.-M.}\ \bibnamefont {Ruan}},\ }\href
  {\doibase 10.1007/JHEP10(2018)011} {\bibfield  {journal} {\bibinfo  {journal}
  {JHEP}\ }\textbf {\bibinfo {volume} {10}},\ \bibinfo {pages} {011} (\bibinfo
  {year} {2018})},\ \Eprint {http://arxiv.org/abs/1807.07677} {arXiv:1807.07677
  [hep-th]} \BibitemShut {NoStop}%
%%CITATION = ARXIV:1807.07677;%%
\bibitem [{\citenamefont {Bhattacharyya}\ \emph
  {et~al.}(2018{\natexlab{b}})\citenamefont {Bhattacharyya}, \citenamefont
  {Shekar},\ and\ \citenamefont {Sinha}}]{Bhattacharyya:2018bbv}%
  \BibitemOpen
  \bibfield  {author} {\bibinfo {author} {\bibfnamefont {A.}~\bibnamefont
  {Bhattacharyya}}, \bibinfo {author} {\bibfnamefont {A.}~\bibnamefont
  {Shekar}}, \ and\ \bibinfo {author} {\bibfnamefont {A.}~\bibnamefont
  {Sinha}},\ }\href {\doibase 10.1007/JHEP10(2018)140} {\bibfield  {journal}
  {\bibinfo  {journal} {JHEP}\ }\textbf {\bibinfo {volume} {10}},\ \bibinfo
  {pages} {140} (\bibinfo {year} {2018}{\natexlab{b}})},\ \Eprint
  {http://arxiv.org/abs/1808.03105} {arXiv:1808.03105 [hep-th]} \BibitemShut
  {NoStop}%
%%CITATION = ARXIV:1808.03105;%%
\bibitem [{\citenamefont {Chapman}\ \emph
  {et~al.}(2018{\natexlab{e}})\citenamefont {Chapman}, \citenamefont {Eisert},
  \citenamefont {Hackl}, \citenamefont {Heller}, \citenamefont {Jefferson},
  \citenamefont {Marrochio},\ and\ \citenamefont {Myers}}]{Chapman:2018hou}%
  \BibitemOpen
  \bibfield  {author} {\bibinfo {author} {\bibfnamefont {S.}~\bibnamefont
  {Chapman}}, \bibinfo {author} {\bibfnamefont {J.}~\bibnamefont {Eisert}},
  \bibinfo {author} {\bibfnamefont {L.}~\bibnamefont {Hackl}}, \bibinfo
  {author} {\bibfnamefont {M.~P.}\ \bibnamefont {Heller}}, \bibinfo {author}
  {\bibfnamefont {R.}~\bibnamefont {Jefferson}}, \bibinfo {author}
  {\bibfnamefont {H.}~\bibnamefont {Marrochio}}, \ and\ \bibinfo {author}
  {\bibfnamefont {R.~C.}\ \bibnamefont {Myers}},\ }\href@noop {} {\  (\bibinfo
  {year} {2018}{\natexlab{e}})},\ \Eprint {http://arxiv.org/abs/1810.05151}
  {arXiv:1810.05151 [hep-th]} \BibitemShut {NoStop}%
%%CITATION = ARXIV:1810.05151;%%
\bibitem [{\citenamefont {Alves}\ and\ \citenamefont
  {Camilo}(2018)}]{Alves:2018qfv}%
  \BibitemOpen
  \bibfield  {author} {\bibinfo {author} {\bibfnamefont {D.~W.~F.}\
  \bibnamefont {Alves}}\ and\ \bibinfo {author} {\bibfnamefont
  {G.}~\bibnamefont {Camilo}},\ }\href {\doibase 10.1007/JHEP06(2018)029}
  {\bibfield  {journal} {\bibinfo  {journal} {JHEP}\ }\textbf {\bibinfo
  {volume} {06}},\ \bibinfo {pages} {029} (\bibinfo {year} {2018})},\ \Eprint
  {http://arxiv.org/abs/1804.00107} {arXiv:1804.00107 [hep-th]} \BibitemShut
  {NoStop}%
%%CITATION = ARXIV:1804.00107;%%
\bibitem [{\citenamefont {Camargo}\ \emph {et~al.}(2018)\citenamefont
  {Camargo}, \citenamefont {Caputa}, \citenamefont {Das}, \citenamefont
  {Heller},\ and\ \citenamefont {Jefferson}}]{Camargo:2018eof}%
  \BibitemOpen
  \bibfield  {author} {\bibinfo {author} {\bibfnamefont {H.~A.}\ \bibnamefont
  {Camargo}}, \bibinfo {author} {\bibfnamefont {P.}~\bibnamefont {Caputa}},
  \bibinfo {author} {\bibfnamefont {D.}~\bibnamefont {Das}}, \bibinfo {author}
  {\bibfnamefont {M.~P.}\ \bibnamefont {Heller}}, \ and\ \bibinfo {author}
  {\bibfnamefont {R.}~\bibnamefont {Jefferson}},\ }\href@noop {} {\  (\bibinfo
  {year} {2018})},\ \Eprint {http://arxiv.org/abs/1807.07075} {arXiv:1807.07075
  [hep-th]} \BibitemShut {NoStop}%
%%CITATION = ARXIV:1807.07075;%%
\bibitem [{\citenamefont {Ali}\ \emph {et~al.}(2018{\natexlab{a}})\citenamefont
  {Ali}, \citenamefont {Bhattacharyya}, \citenamefont {Shajidul~Haque},
  \citenamefont {Kim},\ and\ \citenamefont {Moynihan}}]{Ali:2018fcz}%
  \BibitemOpen
  \bibfield  {author} {\bibinfo {author} {\bibfnamefont {T.}~\bibnamefont
  {Ali}}, \bibinfo {author} {\bibfnamefont {A.}~\bibnamefont {Bhattacharyya}},
  \bibinfo {author} {\bibfnamefont {S.}~\bibnamefont {Shajidul~Haque}},
  \bibinfo {author} {\bibfnamefont {E.~H.}\ \bibnamefont {Kim}}, \ and\
  \bibinfo {author} {\bibfnamefont {N.}~\bibnamefont {Moynihan}},\ }\href@noop
  {} {\  (\bibinfo {year} {2018}{\natexlab{a}})},\ \Eprint
  {http://arxiv.org/abs/1810.02734} {arXiv:1810.02734 [hep-th]} \BibitemShut
  {NoStop}%
%%CITATION = ARXIV:1810.02734;%%
\bibitem [{\citenamefont {Jiang}\ and\ \citenamefont
  {Liu}(2019)}]{Jiang:2018nzg}%
  \BibitemOpen
  \bibfield  {author} {\bibinfo {author} {\bibfnamefont {J.}~\bibnamefont
  {Jiang}}\ and\ \bibinfo {author} {\bibfnamefont {X.}~\bibnamefont {Liu}},\
  }\href {\doibase 10.1103/PhysRevD.99.026011} {\bibfield  {journal} {\bibinfo
  {journal} {Phys. Rev.}\ }\textbf {\bibinfo {volume} {D99}},\ \bibinfo {pages}
  {026011} (\bibinfo {year} {2019})},\ \Eprint
  {http://arxiv.org/abs/1812.00193} {arXiv:1812.00193 [hep-th]} \BibitemShut
  {NoStop}%
%%CITATION = ARXIV:1812.00193;%%
\bibitem [{\citenamefont {Roberts}\ and\ \citenamefont
  {Yoshida}(2017)}]{Roberts:2016hpo}%
  \BibitemOpen
  \bibfield  {author} {\bibinfo {author} {\bibfnamefont {D.~A.}\ \bibnamefont
  {Roberts}}\ and\ \bibinfo {author} {\bibfnamefont {B.}~\bibnamefont
  {Yoshida}},\ }\href {\doibase 10.1007/JHEP04(2017)121} {\bibfield  {journal}
  {\bibinfo  {journal} {JHEP}\ }\textbf {\bibinfo {volume} {04}},\ \bibinfo
  {pages} {121} (\bibinfo {year} {2017})},\ \Eprint
  {http://arxiv.org/abs/1610.04903} {arXiv:1610.04903 [quant-ph]} \BibitemShut
  {NoStop}%
%%CITATION = ARXIV:1610.04903;%%
\bibitem [{\citenamefont {Hashimoto}\ \emph {et~al.}(2017)\citenamefont
  {Hashimoto}, \citenamefont {Iizuka},\ and\ \citenamefont
  {Sugishita}}]{Hashimoto:2017fga}%
  \BibitemOpen
  \bibfield  {author} {\bibinfo {author} {\bibfnamefont {K.}~\bibnamefont
  {Hashimoto}}, \bibinfo {author} {\bibfnamefont {N.}~\bibnamefont {Iizuka}}, \
  and\ \bibinfo {author} {\bibfnamefont {S.}~\bibnamefont {Sugishita}},\ }\href
  {\doibase 10.1103/PhysRevD.96.126001} {\bibfield  {journal} {\bibinfo
  {journal} {Phys. Rev.}\ }\textbf {\bibinfo {volume} {D96}},\ \bibinfo {pages}
  {126001} (\bibinfo {year} {2017})},\ \Eprint
  {http://arxiv.org/abs/1707.03840} {arXiv:1707.03840 [hep-th]} \BibitemShut
  {NoStop}%
%%CITATION = ARXIV:1707.03840;%%
\bibitem [{\citenamefont {Czech}(2017)}]{Czech:2017ryf}%
  \BibitemOpen
  \bibfield  {author} {\bibinfo {author} {\bibfnamefont {B.}~\bibnamefont
  {Czech}},\ }\href@noop {} {\  (\bibinfo {year} {2017})},\ \Eprint
  {http://arxiv.org/abs/1706.00965} {arXiv:1706.00965 [hep-th]} \BibitemShut
  {NoStop}%
%%CITATION = ARXIV:1706.00965;%%
\bibitem [{\citenamefont {Reynolds}\ and\ \citenamefont
  {Ross}(2017{\natexlab{b}})}]{Reynolds:2017lwq}%
  \BibitemOpen
  \bibfield  {author} {\bibinfo {author} {\bibfnamefont {A.}~\bibnamefont
  {Reynolds}}\ and\ \bibinfo {author} {\bibfnamefont {S.~F.}\ \bibnamefont
  {Ross}},\ }\href {\doibase 10.1088/1361-6382/aa8122} {\bibfield  {journal}
  {\bibinfo  {journal} {Class. Quant. Grav.}\ }\textbf {\bibinfo {volume}
  {34}},\ \bibinfo {pages} {175013} (\bibinfo {year} {2017}{\natexlab{b}})},\
  \Eprint {http://arxiv.org/abs/1706.03788} {arXiv:1706.03788 [hep-th]}
  \BibitemShut {NoStop}%
%%CITATION = ARXIV:1706.03788;%%
\bibitem [{\citenamefont {Magan}(2018)}]{Magan:2018nmu}%
  \BibitemOpen
  \bibfield  {author} {\bibinfo {author} {\bibfnamefont {J.~M.}\ \bibnamefont
  {Magan}},\ }\href {\doibase 10.1007/JHEP09(2018)043} {\bibfield  {journal}
  {\bibinfo  {journal} {JHEP}\ }\textbf {\bibinfo {volume} {09}},\ \bibinfo
  {pages} {043} (\bibinfo {year} {2018})},\ \Eprint
  {http://arxiv.org/abs/1805.05839} {arXiv:1805.05839 [hep-th]} \BibitemShut
  {NoStop}%
%%CITATION = ARXIV:1805.05839;%%
\bibitem [{\citenamefont {Caputa}\ and\ \citenamefont
  {Magan}(2018)}]{Caputa:2018kdj}%
  \BibitemOpen
  \bibfield  {author} {\bibinfo {author} {\bibfnamefont {P.}~\bibnamefont
  {Caputa}}\ and\ \bibinfo {author} {\bibfnamefont {J.~M.}\ \bibnamefont
  {Magan}},\ }\href@noop {} {\  (\bibinfo {year} {2018})},\ \Eprint
  {http://arxiv.org/abs/1807.04422} {arXiv:1807.04422 [hep-th]} \BibitemShut
  {NoStop}%
%%CITATION = ARXIV:1807.04422;%%
\bibitem [{\citenamefont {Balasubramanian}\ \emph {et~al.}(2018)\citenamefont
  {Balasubramanian}, \citenamefont {DeCross}, \citenamefont {Kar},\ and\
  \citenamefont {Parrikar}}]{Balasubramanian:2018hsu}%
  \BibitemOpen
  \bibfield  {author} {\bibinfo {author} {\bibfnamefont {V.}~\bibnamefont
  {Balasubramanian}}, \bibinfo {author} {\bibfnamefont {M.}~\bibnamefont
  {DeCross}}, \bibinfo {author} {\bibfnamefont {A.}~\bibnamefont {Kar}}, \ and\
  \bibinfo {author} {\bibfnamefont {O.}~\bibnamefont {Parrikar}},\ }\href@noop
  {} {\  (\bibinfo {year} {2018})},\ \Eprint {http://arxiv.org/abs/1811.04085}
  {arXiv:1811.04085 [hep-th]} \BibitemShut {NoStop}%
%%CITATION = ARXIV:1811.04085;%%
\bibitem [{\citenamefont {Jackiw}(1985)}]{Jackiw:1984je}%
  \BibitemOpen
  \bibfield  {author} {\bibinfo {author} {\bibfnamefont {R.}~\bibnamefont
  {Jackiw}},\ }\bibfield  {booktitle} {\emph {\bibinfo {booktitle} {{1984
  Meeting of the Division of Particles and Fields of the APS Santa Fe, New
  Mexico, October 31-November 3, 1984}}},\ }\href {\doibase
  10.1016/0550-3213(85)90448-1} {\bibfield  {journal} {\bibinfo  {journal}
  {Nucl. Phys.}\ }\textbf {\bibinfo {volume} {B252}},\ \bibinfo {pages} {343}
  (\bibinfo {year} {1985})}\BibitemShut {NoStop}%
%%CITATION = NUPHA,B252,343;%%
\bibitem [{\citenamefont {Shimaji}\ \emph {et~al.}(2018)\citenamefont
  {Shimaji}, \citenamefont {Takayanagi},\ and\ \citenamefont
  {Wei}}]{Shimaji:2018czt}%
  \BibitemOpen
  \bibfield  {author} {\bibinfo {author} {\bibfnamefont {T.}~\bibnamefont
  {Shimaji}}, \bibinfo {author} {\bibfnamefont {T.}~\bibnamefont {Takayanagi}},
  \ and\ \bibinfo {author} {\bibfnamefont {Z.}~\bibnamefont {Wei}},\
  }\href@noop {} {\  (\bibinfo {year} {2018})},\ \Eprint
  {http://arxiv.org/abs/1812.01176} {arXiv:1812.01176 [hep-th]} \BibitemShut
  {NoStop}%
%%CITATION = ARXIV:1812.01176;%%
\bibitem [{\citenamefont {Ali}\ \emph {et~al.}(2018{\natexlab{b}})\citenamefont
  {Ali}, \citenamefont {Bhattacharyya}, \citenamefont {Shajidul~Haque},
  \citenamefont {Kim},\ and\ \citenamefont {Moynihan}}]{Ali:2018aon}%
  \BibitemOpen
  \bibfield  {author} {\bibinfo {author} {\bibfnamefont {T.}~\bibnamefont
  {Ali}}, \bibinfo {author} {\bibfnamefont {A.}~\bibnamefont {Bhattacharyya}},
  \bibinfo {author} {\bibfnamefont {S.}~\bibnamefont {Shajidul~Haque}},
  \bibinfo {author} {\bibfnamefont {E.~H.}\ \bibnamefont {Kim}}, \ and\
  \bibinfo {author} {\bibfnamefont {N.}~\bibnamefont {Moynihan}},\ }\href@noop
  {} {\  (\bibinfo {year} {2018}{\natexlab{b}})},\ \Eprint
  {http://arxiv.org/abs/1811.05985} {arXiv:1811.05985 [hep-th]} \BibitemShut
  {NoStop}%
%%CITATION = ARXIV:1811.05985;%%
\bibitem [{\citenamefont {Liu}\ \emph {et~al.}(2019)\citenamefont {Liu},
  \citenamefont {Lundgren}, \citenamefont {Titum}, \citenamefont {Garrison},\
  and\ \citenamefont {Gorshkov}}]{Liu:2019aji}%
  \BibitemOpen
  \bibfield  {author} {\bibinfo {author} {\bibfnamefont {F.}~\bibnamefont
  {Liu}}, \bibinfo {author} {\bibfnamefont {R.}~\bibnamefont {Lundgren}},
  \bibinfo {author} {\bibfnamefont {P.}~\bibnamefont {Titum}}, \bibinfo
  {author} {\bibfnamefont {J.~R.}\ \bibnamefont {Garrison}}, \ and\ \bibinfo
  {author} {\bibfnamefont {A.~V.}\ \bibnamefont {Gorshkov}},\ }\href@noop {} {\
   (\bibinfo {year} {2019})},\ \Eprint {http://arxiv.org/abs/1902.10720}
  {arXiv:1902.10720 [quant-ph]} \BibitemShut {NoStop}%
%%CITATION = ARXIV:1902.10720;%%
\bibitem [{\citenamefont {Nielsen}\ \emph {et~al.}(2006)\citenamefont
  {Nielsen}, \citenamefont {Dowling}, \citenamefont {Gu},\ and\ \citenamefont
  {Doherty}}]{nielsen2006quantum}%
  \BibitemOpen
  \bibfield  {author} {\bibinfo {author} {\bibfnamefont {M.~A.}\ \bibnamefont
  {Nielsen}}, \bibinfo {author} {\bibfnamefont {M.~R.}\ \bibnamefont
  {Dowling}}, \bibinfo {author} {\bibfnamefont {M.}~\bibnamefont {Gu}}, \ and\
  \bibinfo {author} {\bibfnamefont {A.~C.}\ \bibnamefont {Doherty}},\
  }\href@noop {} {\bibfield  {journal} {\bibinfo  {journal} {Science}\ }\textbf
  {\bibinfo {volume} {311}},\ \bibinfo {pages} {1133} (\bibinfo {year}
  {2006})}\BibitemShut {NoStop}%
\bibitem [{\citenamefont {Dowling}\ and\ \citenamefont
  {Nielsen}(2008)}]{nielsen2008}%
  \BibitemOpen
  \bibfield  {author} {\bibinfo {author} {\bibfnamefont {M.~R.}\ \bibnamefont
  {Dowling}}\ and\ \bibinfo {author} {\bibfnamefont {M.~A.}\ \bibnamefont
  {Nielsen}},\ }\href@noop {} {\bibfield  {journal} {\bibinfo  {journal}
  {Quantum Info. Comput.}\ }\textbf {\bibinfo {volume} {8}},\ \bibinfo {pages}
  {861} (\bibinfo {year} {2008})}\BibitemShut {NoStop}%
\bibitem [{\citenamefont {Nielsen}(2006)}]{Nielsen:2006}%
  \BibitemOpen
  \bibfield  {author} {\bibinfo {author} {\bibfnamefont {M.~A.}\ \bibnamefont
  {Nielsen}},\ }\href {http://dl.acm.org/citation.cfm?id=2011686.2011688}
  {\bibfield  {journal} {\bibinfo  {journal} {Quantum Info. Comput.}\ }\textbf
  {\bibinfo {volume} {6}},\ \bibinfo {pages} {213} (\bibinfo {year}
  {2006})}\BibitemShut {NoStop}%
\bibitem [{Note1()}]{Note1}%
  \BibitemOpen
  \bibinfo {note} {See Supplemental Material-A for details where the references
  \cite
  {foot01,Avis:1977yn,Burgess:1984ti,Cotabreveescu:1999em,Cotabreveescu:1999em,Fitzpatrick:2011jn,kaplan2013lectures,Terashima:2017gmc,Ammon:2015wua,ElShowk:2011ag,Fitzpatrick:2011jn,kaplan2013lectures,Terashima:2017gmc,Terashima:2017gmc,long-paper,kaplan2013lectures,BottaCantcheff:2015sav,
  Marolf:2017kvq, BottaCantcheff:2019apr} are included. \label
  {SuppMa}}\BibitemShut {NoStop}%
\bibitem [{\citenamefont {Bizon}\ and\ \citenamefont
  {Rostworowski}(2011)}]{Bizon:2011gg}%
  \BibitemOpen
  \bibfield  {author} {\bibinfo {author} {\bibfnamefont {P.}~\bibnamefont
  {Bizon}}\ and\ \bibinfo {author} {\bibfnamefont {A.}~\bibnamefont
  {Rostworowski}},\ }\href {\doibase 10.1103/PhysRevLett.107.031102} {\bibfield
   {journal} {\bibinfo  {journal} {Phys. Rev. Lett.}\ }\textbf {\bibinfo
  {volume} {107}},\ \bibinfo {pages} {031102} (\bibinfo {year} {2011})},\
  \Eprint {http://arxiv.org/abs/1104.3702} {arXiv:1104.3702 [gr-qc]}
  \BibitemShut {NoStop}%
%%CITATION = ARXIV:1104.3702;%%
\bibitem [{\citenamefont {Buchel}\ \emph {et~al.}(2012)\citenamefont {Buchel},
  \citenamefont {Lehner},\ and\ \citenamefont {Liebling}}]{Buchel:2012uh}%
  \BibitemOpen
  \bibfield  {author} {\bibinfo {author} {\bibfnamefont {A.}~\bibnamefont
  {Buchel}}, \bibinfo {author} {\bibfnamefont {L.}~\bibnamefont {Lehner}}, \
  and\ \bibinfo {author} {\bibfnamefont {S.~L.}\ \bibnamefont {Liebling}},\
  }\href {\doibase 10.1103/PhysRevD.86.123011} {\bibfield  {journal} {\bibinfo
  {journal} {Phys. Rev.}\ }\textbf {\bibinfo {volume} {D86}},\ \bibinfo {pages}
  {123011} (\bibinfo {year} {2012})},\ \Eprint {http://arxiv.org/abs/1210.0890}
  {arXiv:1210.0890 [gr-qc]} \BibitemShut {NoStop}%
%%CITATION = ARXIV:1210.0890;%%
\bibitem [{\citenamefont {Buchel}\ \emph {et~al.}(2013)\citenamefont {Buchel},
  \citenamefont {Liebling},\ and\ \citenamefont {Lehner}}]{Buchel:2013uba}%
  \BibitemOpen
  \bibfield  {author} {\bibinfo {author} {\bibfnamefont {A.}~\bibnamefont
  {Buchel}}, \bibinfo {author} {\bibfnamefont {S.~L.}\ \bibnamefont
  {Liebling}}, \ and\ \bibinfo {author} {\bibfnamefont {L.}~\bibnamefont
  {Lehner}},\ }\href {\doibase 10.1103/PhysRevD.87.123006} {\bibfield
  {journal} {\bibinfo  {journal} {Phys. Rev.}\ }\textbf {\bibinfo {volume}
  {D87}},\ \bibinfo {pages} {123006} (\bibinfo {year} {2013})},\ \Eprint
  {http://arxiv.org/abs/1304.4166} {arXiv:1304.4166 [gr-qc]} \BibitemShut
  {NoStop}%
%%CITATION = ARXIV:1304.4166;%%
\bibitem [{foo({\natexlab{a}})}]{foot04}%
  \BibitemOpen
  \href@noop {} {\emph {\bibinfo {title} {The scalar field action has a
  well-defined variational principle without any additional surface terms, and
  so we assume no surface terms are added for the scalar.}}}\BibitemShut
  {Stop}%
\bibitem [{Note2()}]{Note2}%
  \BibitemOpen
  \bibinfo {note} {See Supplemental Material-B for more details where the
  references \cite {foot02,PhysRevLett.28.1082, PhysRevD.15.2752,foot03} are
  included.}\BibitemShut {Stop}%
\bibitem [{\citenamefont {Chapman}\ \emph
  {et~al.}(2018{\natexlab{f}})\citenamefont {Chapman}, \citenamefont
  {Marrochio},\ and\ \citenamefont {Myers}}]{Vad2}%
  \BibitemOpen
  \bibfield  {author} {\bibinfo {author} {\bibfnamefont {S.}~\bibnamefont
  {Chapman}}, \bibinfo {author} {\bibfnamefont {H.}~\bibnamefont {Marrochio}},
  \ and\ \bibinfo {author} {\bibfnamefont {R.~C.}\ \bibnamefont {Myers}},\
  }\href {\doibase 10.1007/JHEP06(2018)114} {\bibfield  {journal} {\bibinfo
  {journal} {JHEP}\ }\textbf {\bibinfo {volume} {06}},\ \bibinfo {pages} {114}
  (\bibinfo {year} {2018}{\natexlab{f}})},\ \Eprint
  {http://arxiv.org/abs/1805.07262} {arXiv:1805.07262 [hep-th]} \BibitemShut
  {NoStop}%
%%CITATION = ARXIV:1805.07262;%%
\bibitem [{\citenamefont {Bernamonti}\ \emph {et~al.}(2020)\citenamefont
  {Bernamonti}, \citenamefont {Galli}, \citenamefont {Hernandez}, \citenamefont
  {Myers}, \citenamefont {Ruan},\ and\ \citenamefont {Sim\'on}}]{long-paper}%
  \BibitemOpen
  \bibfield  {author} {\bibinfo {author} {\bibfnamefont {A.}~\bibnamefont
  {Bernamonti}}, \bibinfo {author} {\bibfnamefont {F.}~\bibnamefont {Galli}},
  \bibinfo {author} {\bibfnamefont {J.}~\bibnamefont {Hernandez}}, \bibinfo
  {author} {\bibfnamefont {R.~C.}\ \bibnamefont {Myers}}, \bibinfo {author}
  {\bibfnamefont {S.-M.}\ \bibnamefont {Ruan}}, \ and\ \bibinfo {author}
  {\bibfnamefont {J.}~\bibnamefont {Sim\'on}},\ }\href@noop {} {\  (\bibinfo
  {year} {2020})},\ \Eprint {http://arxiv.org/abs/2002.05779} {arXiv:2002.05779
  [hep-th]} \BibitemShut {NoStop}%
%%CITATION = ARXIV:2002.05779;%%
\bibitem [{\citenamefont {Miyaji}\ and\ \citenamefont
  {Takayanagi}(2015)}]{Miyaji:2015yva}%
  \BibitemOpen
  \bibfield  {author} {\bibinfo {author} {\bibfnamefont {M.}~\bibnamefont
  {Miyaji}}\ and\ \bibinfo {author} {\bibfnamefont {T.}~\bibnamefont
  {Takayanagi}},\ }\href {\doibase 10.1093/ptep/ptv089} {\bibfield  {journal}
  {\bibinfo  {journal} {PTEP}\ }\textbf {\bibinfo {volume} {2015}},\ \bibinfo
  {pages} {073B03} (\bibinfo {year} {2015})},\ \Eprint
  {http://arxiv.org/abs/1503.03542} {arXiv:1503.03542 [hep-th]} \BibitemShut
  {NoStop}%
%%CITATION = ARXIV:1503.03542;%%
\bibitem [{foo({\natexlab{b}})}]{foot01}%
  \BibitemOpen
  \href@noop {} {\emph {\bibinfo {title} {Hence the standard normalization for
  a scalar in QFT calculations is achieved by defining as $\Psi =
  \Phi/\sqrt{16\pi\GN}$.}}}\BibitemShut {Stop}%
\bibitem [{\citenamefont {Avis}\ \emph {et~al.}(1978)\citenamefont {Avis},
  \citenamefont {Isham},\ and\ \citenamefont {Storey}}]{Avis:1977yn}%
  \BibitemOpen
  \bibfield  {author} {\bibinfo {author} {\bibfnamefont {S.~J.}\ \bibnamefont
  {Avis}}, \bibinfo {author} {\bibfnamefont {C.~J.}\ \bibnamefont {Isham}}, \
  and\ \bibinfo {author} {\bibfnamefont {D.}~\bibnamefont {Storey}},\ }\href
  {\doibase 10.1103/PhysRevD.18.3565} {\bibfield  {journal} {\bibinfo
  {journal} {Phys. Rev.}\ }\textbf {\bibinfo {volume} {D18}},\ \bibinfo {pages}
  {3565} (\bibinfo {year} {1978})}\BibitemShut {NoStop}%
%%CITATION = PHRVA,D18,3565;%%
\bibitem [{\citenamefont {Burgess}\ and\ \citenamefont
  {Lutken}(1985)}]{Burgess:1984ti}%
  \BibitemOpen
  \bibfield  {author} {\bibinfo {author} {\bibfnamefont {C.~P.}\ \bibnamefont
  {Burgess}}\ and\ \bibinfo {author} {\bibfnamefont {C.~A.}\ \bibnamefont
  {Lutken}},\ }\href {\doibase 10.1016/0370-2693(85)91415-7} {\bibfield
  {journal} {\bibinfo  {journal} {Phys. Lett.}\ }\textbf {\bibinfo {volume}
  {153B}},\ \bibinfo {pages} {137} (\bibinfo {year} {1985})}\BibitemShut
  {NoStop}%
%%CITATION = PHLTA,153B,137;%%
\bibitem [{\citenamefont {Cot\ifmmode~\u{a}\else
  \u{a}\fi{}escu}(1999)}]{Cotabreveescu:1999em}%
  \BibitemOpen
  \bibfield  {author} {\bibinfo {author} {\bibfnamefont {I.~I.}\ \bibnamefont
  {Cot\ifmmode~\u{a}\else \u{a}\fi{}escu}},\ }\href {\doibase
  10.1103/PhysRevD.60.107504} {\bibfield  {journal} {\bibinfo  {journal} {Phys.
  Rev. D}\ }\textbf {\bibinfo {volume} {60}},\ \bibinfo {pages} {107504}
  (\bibinfo {year} {1999})}\BibitemShut {NoStop}%
\bibitem [{\citenamefont {Fitzpatrick}\ and\ \citenamefont
  {Kaplan}(2011)}]{Fitzpatrick:2011jn}%
  \BibitemOpen
  \bibfield  {author} {\bibinfo {author} {\bibfnamefont {A.~L.}\ \bibnamefont
  {Fitzpatrick}}\ and\ \bibinfo {author} {\bibfnamefont {J.}~\bibnamefont
  {Kaplan}},\ }\href@noop {} {\  (\bibinfo {year} {2011})},\ \Eprint
  {http://arxiv.org/abs/1104.2597} {arXiv:1104.2597 [hep-th]} \BibitemShut
  {NoStop}%
%%CITATION = ARXIV:1104.2597;%%
\bibitem [{\citenamefont {Kaplan}(2013)}]{kaplan2013lectures}%
  \BibitemOpen
  \bibfield  {author} {\bibinfo {author} {\bibfnamefont {J.}~\bibnamefont
  {Kaplan}},\ }\href@noop {} {\enquote {\bibinfo {title} {Lectures on
  \text{AdS/CFT} from the bottom up},}\ } (\bibinfo {year} {2013})\BibitemShut
  {NoStop}%
\bibitem [{\citenamefont {Terashima}(2018)}]{Terashima:2017gmc}%
  \BibitemOpen
  \bibfield  {author} {\bibinfo {author} {\bibfnamefont {S.}~\bibnamefont
  {Terashima}},\ }\href {\doibase 10.1007/JHEP02(2018)019} {\bibfield
  {journal} {\bibinfo  {journal} {JHEP}\ }\textbf {\bibinfo {volume} {02}},\
  \bibinfo {pages} {019} (\bibinfo {year} {2018})},\ \Eprint
  {http://arxiv.org/abs/1710.07298} {arXiv:1710.07298 [hep-th]} \BibitemShut
  {NoStop}%
%%CITATION = ARXIV:1710.07298;%%
\bibitem [{\citenamefont {Ammon}\ and\ \citenamefont
  {Erdmenger}(2015)}]{Ammon:2015wua}%
  \BibitemOpen
  \bibfield  {author} {\bibinfo {author} {\bibfnamefont {M.}~\bibnamefont
  {Ammon}}\ and\ \bibinfo {author} {\bibfnamefont {J.}~\bibnamefont
  {Erdmenger}},\ }\href@noop {} {\emph {\bibinfo {title} {{Gauge/gravity
  duality}}}}\ (\bibinfo  {publisher} {Cambridge University Press},\ \bibinfo
  {address} {Cambridge},\ \bibinfo {year} {2015})\BibitemShut {NoStop}%
\bibitem [{\citenamefont {El-Showk}\ and\ \citenamefont
  {Papadodimas}(2012)}]{ElShowk:2011ag}%
  \BibitemOpen
  \bibfield  {author} {\bibinfo {author} {\bibfnamefont {S.}~\bibnamefont
  {El-Showk}}\ and\ \bibinfo {author} {\bibfnamefont {K.}~\bibnamefont
  {Papadodimas}},\ }\href {\doibase 10.1007/JHEP10(2012)106} {\bibfield
  {journal} {\bibinfo  {journal} {JHEP}\ }\textbf {\bibinfo {volume} {10}},\
  \bibinfo {pages} {106} (\bibinfo {year} {2012})},\ \Eprint
  {http://arxiv.org/abs/1101.4163} {arXiv:1101.4163 [hep-th]} \BibitemShut
  {NoStop}%
%%CITATION = ARXIV:1101.4163;%%
\bibitem [{\citenamefont {Botta-Cantcheff}\ \emph {et~al.}(2016)\citenamefont
  {Botta-Cantcheff}, \citenamefont {Mart\'inez},\ and\ \citenamefont
  {Silva}}]{BottaCantcheff:2015sav}%
  \BibitemOpen
  \bibfield  {author} {\bibinfo {author} {\bibfnamefont {M.}~\bibnamefont
  {Botta-Cantcheff}}, \bibinfo {author} {\bibfnamefont {P.}~\bibnamefont
  {Mart\'inez}}, \ and\ \bibinfo {author} {\bibfnamefont {G.~A.}\ \bibnamefont
  {Silva}},\ }\href {\doibase 10.1007/JHEP02(2016)171} {\bibfield  {journal}
  {\bibinfo  {journal} {JHEP}\ }\textbf {\bibinfo {volume} {02}},\ \bibinfo
  {pages} {171} (\bibinfo {year} {2016})},\ \Eprint
  {http://arxiv.org/abs/1512.07850} {arXiv:1512.07850 [hep-th]} \BibitemShut
  {NoStop}%
%%CITATION = ARXIV:1512.07850;%%
\bibitem [{\citenamefont {Marolf}\ \emph {et~al.}(2018)\citenamefont {Marolf},
  \citenamefont {Parrikar}, \citenamefont {Rabideau}, \citenamefont
  {Izadi~Rad},\ and\ \citenamefont {Van~Raamsdonk}}]{Marolf:2017kvq}%
  \BibitemOpen
  \bibfield  {author} {\bibinfo {author} {\bibfnamefont {D.}~\bibnamefont
  {Marolf}}, \bibinfo {author} {\bibfnamefont {O.}~\bibnamefont {Parrikar}},
  \bibinfo {author} {\bibfnamefont {C.}~\bibnamefont {Rabideau}}, \bibinfo
  {author} {\bibfnamefont {A.}~\bibnamefont {Izadi~Rad}}, \ and\ \bibinfo
  {author} {\bibfnamefont {M.}~\bibnamefont {Van~Raamsdonk}},\ }\href {\doibase
  10.1007/JHEP06(2018)077} {\bibfield  {journal} {\bibinfo  {journal} {JHEP}\
  }\textbf {\bibinfo {volume} {06}},\ \bibinfo {pages} {077} (\bibinfo {year}
  {2018})},\ \Eprint {http://arxiv.org/abs/1709.10101} {arXiv:1709.10101
  [hep-th]} \BibitemShut {NoStop}%
%%CITATION = ARXIV:1709.10101;%%
\bibitem [{\citenamefont {Botta-Cantcheff}\ \emph {et~al.}(2019)\citenamefont
  {Botta-Cantcheff}, \citenamefont {Mart\'{i}nez},\ and\ \citenamefont
  {Silva}}]{BottaCantcheff:2019apr}%
  \BibitemOpen
  \bibfield  {author} {\bibinfo {author} {\bibfnamefont {M.}~\bibnamefont
  {Botta-Cantcheff}}, \bibinfo {author} {\bibfnamefont {P.~J.}\ \bibnamefont
  {Mart\'{i}nez}}, \ and\ \bibinfo {author} {\bibfnamefont {G.~A.}\
  \bibnamefont {Silva}},\ }\href@noop {} {\  (\bibinfo {year} {2019})},\
  \Eprint {http://arxiv.org/abs/1901.00505} {arXiv:1901.00505 [hep-th]}
  \BibitemShut {NoStop}%
%%CITATION = ARXIV:1901.00505;%%
\bibitem [{\citenamefont {York}(1972)}]{PhysRevLett.28.1082}%
  \BibitemOpen
  \bibfield  {author} {\bibinfo {author} {\bibfnamefont {J.~W.}\ \bibnamefont
  {York}},\ }\href {\doibase 10.1103/PhysRevLett.28.1082} {\bibfield  {journal}
  {\bibinfo  {journal} {Phys. Rev. Lett.}\ }\textbf {\bibinfo {volume} {28}},\
  \bibinfo {pages} {1082} (\bibinfo {year} {1972})}\BibitemShut {NoStop}%
\bibitem [{\citenamefont {Gibbons}\ and\ \citenamefont
  {Hawking}(1977)}]{PhysRevD.15.2752}%
  \BibitemOpen
  \bibfield  {author} {\bibinfo {author} {\bibfnamefont {G.~W.}\ \bibnamefont
  {Gibbons}}\ and\ \bibinfo {author} {\bibfnamefont {S.~W.}\ \bibnamefont
  {Hawking}},\ }\href {\doibase 10.1103/PhysRevD.15.2752} {\bibfield  {journal}
  {\bibinfo  {journal} {Phys. Rev. D}\ }\textbf {\bibinfo {volume} {15}},\
  \bibinfo {pages} {2752} (\bibinfo {year} {1977})}\BibitemShut {NoStop}%
\bibitem [{foo({\natexlab{c}})}]{foot02}%
  \BibitemOpen
  \href@noop {} {\emph {\bibinfo {title} {No horizon is expected for a
  perturbative excitation of the AdS spacetime, but technically the possibility
  of opening an event horizon is removed by the boundary condition
  $a_2(t,\rho=0)=0$, which corresponds to the geometry closing off smoothly at
  the origin of the radius.}}}\BibitemShut {Stop}%
\bibitem [{foo({\natexlab{d}})}]{foot03}%
  \BibitemOpen
  \href@noop {} {\emph {\bibinfo {title} {The first equality can be verified
  writing the integrals explicitly in terms of the radial variable and noticing
  that within our perturbative setup, $ds = (d\rho / k_{0}^{\rho})(1-\delta
  k^\rho)$ on the boundary of the deformed WDW patch.}}}\BibitemShut {Stop}%
\end{thebibliography}%

\pagebreak

\appendix

\titlepage

\setcounter{page}{1}

\begin{center}
{\LARGE Supplemental Material}
\end{center}

\section{A. Coherent States in the Bulk and on the Boundary}

The details of the construction of the coherent states studied in the main text are discussed below. We consider a massless real scalar field on the $\text{AdS}_{4}$ spacetime with global Lorentzian coordinates
\begin{equation}\label{vac2}
ds^2_{\mt{AdS}}= \frac{L^2}{\cos^2\! \rho}\( -dt^2+d\rho^2 +\sin^2 \!\rho \, (d\theta^2+\sin^2\theta\,d\phi^2)\)\,,
\end{equation}
where $L$ denotes the radius of $\text{AdS}$ spacetime with dimensionless coordinates $y^\mu=(t,\rho,\theta,\phi)$.
The free scalar field theory is described by the action 
\begin{equation}
I_{\mt{matter}}= -\frac{1}{32 \pi\GN } \int d^{4} y\,\sqrt{-g}\left( g^{\mu\nu}\nabla_\mu \Phi \nabla_\nu \Phi \)\,,
\end{equation}
where we keep the convention adopted in the main text of the letter of including an overall factor $\frac{1}{16\pi \GN}$, which is convenient in solving Einstein's equations for the backreaction \cite{foot01}.

The analysis of solutions to the corresponding Klein-Gordon equation  $\partial_{\mu} \(\sqrt{-g}g^{\mu\nu}\partial_\nu\Phi  \)=0$ yields the discrete spectrum \cite{Avis:1977yn,Burgess:1984ti,Cotabreveescu:1999em} 
\begin{equation}\label{eigenenergy}
\omega_n  = 3 + 2j+\ell\,, 
\end{equation} 
for the eigenfunctions 
\begin{equation}\label{eigenwaves}
u_{n}(y^\mu)
={N}_{j,\ell}\ \sin^\ell\! \rho\,\cos^3\! \rho\ {}_2 F_1\!\(-j,3 +j+\ell; \frac{3}{2}+\ell;\sin^2\!\rho\) \,Y_{\ell{m}} \!\(\theta,\phi\) \,e^{- i \omega_{n} t}\, .
\end{equation}
Here $Y_{\ell{m}} \!\(\theta,\phi\)$ are the standard spherical harmonics on the  two-sphere, and the quantum numbers are collectively denoted as $n=(j,\ell,{m})$. It is worth noting the eigenfrequencies $\omega_n$ are dimensionless because the
time coordinate $t$ in eq.~\reef{vac2} is dimensionless. The normalization of these eigenfunctions $u_{n}$ can be fixed by the inner product on a constant time slice $\Sigma_t$ in the above metric \reef{vac2},
\begin{equation}\label{innerp}
\begin{split}
\langle u_{n}, u_{n'} \rangle &=\frac{-i}{16\pi \GN}\int_{\Sigma_t} d^3 y\, \sqrt{-g}\,g^{00} \( u^\ast_{n}\, {\partial_t} u_{n'}
-{\partial_t}u^\ast_{n} \, u_{n'}\) = \delta_{nn'}=-\langle u^\ast_{n}, u^\ast_{n'} \rangle \,,\qquad 
\langle u_{n}, u^\ast_{n'} \rangle =0\,,\\
\end{split}
\end{equation}
yielding  \cite{Cotabreveescu:1999em}
\begin{equation}
{N}_{j,l}=4\,(-)^j\, \sqrt{ \frac{\pi \GN}{L^{2}}\,\frac{ \Gamma(j+\ell+\frac{3}{2})\,\Gamma(j+\ell+3)}{ \Gamma^2(\ell+\frac{3}{2})\,\Gamma(j+\frac{5}2)\,\Gamma(j+1)}}\ .
\end{equation}
For the spherically symmetric modes with $n=(j,0,0)$, these eigenfunctions \reef{eigenwaves} then reduce to the expressions given in eqs.~\reef{eigenmode} and \reef{mode} (using $Y_{00} \!\(\theta,\phi\)=1/\sqrt{4\pi}$). 

The canonical quantum field operator follows (\eg see  \cite{Cotabreveescu:1999em,Fitzpatrick:2011jn,kaplan2013lectures,Terashima:2017gmc})
\begin{equation}\label{def_scalar_quantization}
\hat{\Phi}(y^\mu) = \sum_{n} \(  u_{n}(y^\mu)\ {a}_{n} +  u^{\ast}_{n}(y^\mu)\ {a}^{\dagger}_{n} \) \,.
\end{equation}
Using the normalization chosen above, the annihilation and creation operators here satisfy the standard commutator $[  { a}_{n}, { a}^\dagger_{n^{'}} ]=  \delta_{nn'}$. Using the inner product \reef{innerp}, one can easily extract the latter from the canonical field operator, \ie
\begin{equation}\label{aadagger}
a_{n}= \langle u_{n}, \hat\Phi \rangle =\frac{-i}{16\pi \GN}\int_{\Sigma_t} d^3 y\, \sqrt{-g}g^{00} \( u_{n}^\ast \overleftrightarrow{\partial_t} \hat\Phi\)\,, \qquad  a_{n}^\dagger= -\langle  u^\ast_{n} ,\hat\Phi\rangle =\frac{i}{16\pi \GN}\int_{\Sigma_t} d^3 y\, \sqrt{-g}g^{00} \( u
_{n} \overleftrightarrow{\partial_t} \hat\Phi\) \,.
\end{equation}
These operators generate a basis of states for the Hilbert space of the quantized scalar in AdS$_4$, \ie
\begin{equation}\label{Fock}
\prod_n \( a_n^\dagger\)^{r_n} \ket{0} \,, \quad r_n \in \mathbb{N}\,.
\end{equation}

In this letter, we focus on coherent states. We introduced the displacement operator for each of the modes 
\begin{equation}\label{displacement}
D(\alpha_{n})= {\alpha_{n}\, a_{n}^{\dagger}-\alpha^{\ast}_{n}\,a_{n}} 
=\frac{i}{16\pi \GN} \int_{\Sigma_t} d^3 y \,\sqrt{-g}g^{00}   \(\alpha_{n}u_n+ \alpha_{n}^\ast u_{n}^\ast \) \overleftrightarrow{\partial_t} \Phi (y^\mu) \,,\\
\end{equation}
with which we can build the coherent state for the $n$-mode with amplitude $\alpha_n$ 
\begin{equation}\label{coherent_state_CFT}
\ket{\alpha_{n}}= e^{D(\alpha_{n})} \ket{0}= e^{-\frac 12 |\alpha_n|^2} e^{\alpha_n a^\dagger_n}  \ket{0}\,,
\end{equation}
\ie $a_n \ket{\alpha_{n}}=\alpha_n\ket{\alpha_{n}}$. In any of these individual states, the expectation value of  the quantum scalar field becomes  
\begin{equation}
\bra{\alpha_n} \hat{\Phi}(y^\mu) \ket{\alpha_n}=\left(\alpha_n\,u_n+\alpha_n^\ast\,u^\ast_n \right)\,.
\end{equation}
That is, we have turned on the $n$-mode with a classical amplitude $\alpha_n$. For the multi-mode coherent states in eq.~\reef{Def_coherent}  considered in the main text, the expectation value is given by a superposition of excitations in the corresponding modes, as in eq.~\reef{class}.

According to the standard AdS/CFT dictionary, \eg \cite{Ammon:2015wua}, the bulk massless scalar $\Phi$ is dual to a marginal boundary operator $\hat\cO$ with conformal dimension $\Delta=3$. Furthermore, at leading order in the large $N$ limit, the quantized scalar field in AdS spacetime is dual to a generalized free field operator  in the boundary CFT described in terms of the same annihilation and creation operators. This, in turn, implies that the CFT contains a Hilbert space equivalent to eq.~\eqref{Fock}  \cite{ElShowk:2011ag,Fitzpatrick:2011jn,kaplan2013lectures,Terashima:2017gmc}. Of course, this is simply a reflection of the AdS/CFT correspondence,  which states we can think of the bulk gravity theory and the boundary CFT as  providing two alternative descriptions of the same space of quantum states. This identification plays an important role in our attempt to bridge the discussions of quantum circuit complexity and holographic complexity. 

Let us construct the generalized free field operator, following \cite{Fitzpatrick:2011jn}. We introduce a UV regulator surface in the AdS geometry \reef{vac2} at 
\begin{equation}
\rho(\epsilon)= \frac{\pi}{2}- \frac{L}{R}\,\epsilon\,.
\end{equation}
After an appropriate scaling of the asymptotic AdS metric,  the metric for the boundary CFT becomes
\begin{equation}\label{cylinder_metric2}
ds^2_{\mt{CFT}} 
= R^2 \( -dt^2 +  d\theta^2+\sin^2\theta\,d\phi^2 \)= -dT^2 + R^2 \(d\theta^2+\sin^2\theta\,d\phi^2\) \,, 
\end{equation}
where $T= Rt$ is a dimensionful boundary time. Hence the boundary CFT is defined on a cylinder $\mathbb{R}\times \mathbb{S}^{2}$, where the radius of the two-sphere is $R$.
With this choice, the CFT operator $\hat\cO$ generates a spectrum of states with energies 
\begin{equation}
{\Omega}_n = \frac{\omega_n }{R} = \frac{3 +2j+\ell }{R} \,, 
\end{equation}
corresponding to excitations of the vacuum by  $\hat\cO$ and its descendants, \ie
\begin{equation}\label{descend}
 s^{\mu_1\mu_2\cdots \mu_\ell}_{\ell  m} P_{\mu_1}P_{\mu_2}\cdots P_{\mu_\ell}\ (P^2)^j %e^{i \Delta t} 
\, \hat\cO\,,
\end{equation}
where $P_{\mu}$ are the momentum generators, and $s^{\mu_1\mu_2\cdots \mu_l}_{lm} $ is a symmetric traceless tensor, \eg see \cite{Terashima:2017gmc,long-paper}. 

The AdS/CFT correspondence allows us to construct the generalized free field operator in the boundary theory from the dual scalar field operator $\hat{\Phi}(y^\mu)$ in eq.~\reef{def_scalar_quantization} as \cite{kaplan2013lectures,Fitzpatrick:2011jn} 
\begin{equation}\label{newop}
\hat \mO (T,\theta,\phi)  =  \sqrt{ \frac{3  \pi L^{2}}{64\,\GN}}\, \lim_{\rho(\epsilon) \rightarrow \frac{\pi}{2}} \frac{ \hat{\Phi}(t,\rho(\epsilon),\theta,\phi)}{\cos^3\!\rho(\epsilon)}=\sum_{n} \( \tilde u_{n}(T,\theta,\phi)\ {a}_{n} +  \tilde u^{\ast}_{n}(T,\theta,\phi)\ {a}^{\dagger}_{n} \) \,,
\end{equation}
where the eigenmodes are now
\begin{equation}\label{eigenwavess}
\tilde u_{n}(T,\theta, \phi)
=\widetilde{N}_{j,\ell}\ Y_{\ell{m}} \!\(\theta,\phi\) \,e^{- i \Omega_{n} T} 
\qquad{\rm with}\quad 
\widetilde{N}_{j,\ell}=(-)^j\sqrt{\frac{4\pi}3\, \frac{ \Gamma(j+\ell+3)\,\Gamma(j+\frac{5}{2})}{\Gamma(j+\ell+\frac{3}{2})\,\Gamma(j+1)} }\ .
\end{equation}
The normalization $\widetilde{N}_{j,\ell}$ is derived in \cite{Fitzpatrick:2011jn} by requiring that the two point function takes the standard form. The creation and annihilation operators can now be extracted from the boundary operator $\hat\mO$ with
\begin{equation}\label{arms}
  a_{n}= \llangle \tilde{u}_{n}, \hat  \mO (T,\theta,\phi)\rrangle  \,, \qquad a_{n}^\dagger= \llangle \hat \mO (T,\theta,\phi),\tilde{u}^\ast_{n} \rrangle \,,
\end{equation}
where we have defined the boundary ``inner product'' satisfying
\begin{equation}\label{inner}
\llangle \tilde{u}_{n}, \tilde{u}_{n'} \rrangle =\frac{i }{ 4\pi R\,  \Omega_n\, \tilde{N}_{j,\ell}^2}\int^{2\pi R}_0\!\! dT \int d^2\Omega  \,\( \tilde{u}^\ast_{n}\, {\partial_T} \tilde{u}_{n'}
-{\partial_T}\tilde{u}^\ast_{n} \, \tilde{u}_{n'}\) = \delta_{nn'}\,,\qquad 
\llangle \tilde{u}_{n}, \tilde{u}^\ast_{n'} \rrangle=0\,.
\end{equation}
Note that we must include an integral over time here because the spatial part of the wavefunctions $\tilde u_n$ is insensitive to the quantum number $j$. A more traditional approach would associate the creation operators to the states created by the boundary operator and its descendants in the Euclidean,  \eg see \cite{Terashima:2017gmc,long-paper}.  The present  construction (in particular eq.~\reef{newop}) makes clear that in both the bulk and boundary theories, we are working with the same Hilbert space \reef{Fock}. 

We can construct the corresponding coherent states \reef{coherent_state_CFT} as excited states in the boundary theory by extracting $a_n$ and $a_n^\dagger$ from the generalized free field \reef{newop} using eq.~\eqref{arms}. The corresponding expectation values become
\begin{equation}
\bra{\alpha_n} \hat \mO (T,\theta,\phi)   \ket{\alpha_n}=\left(\alpha_n\,\tilde u_n+\alpha_n^\ast\,\tilde u^\ast_n \right)\, ,
\end{equation}
and  for the multi-mode coherent states \reef{Def_coherent} in the main text  
\begin{equation}
\bra{\varepsilon \alpha_j} \hat \mO (T,\theta,\phi) \ket{\varepsilon \alpha_j}=\varepsilon
\sum_{\{j\}}\!\big(\alpha_j\,\tilde u_j+\alpha_j^\ast\,\tilde u^\ast_j \big)\equiv \veps\,\mO_{ \mt{cl}}(T,\theta,\phi)\, .\label{expect99}
\end{equation}
Alternatively, using eqs.~\reef{class} and \reef{newop}, we have
\begin{equation}\label{new99}
\mO_{\mt{cl}}(T,\theta,\phi)  =  \sqrt{ \frac{3 \pi L^{2}  }{64\,\GN}}\,\lim_{\rho(\epsilon) \rightarrow \frac{\pi}{2}} \frac{ {\Phi}_\mt{cl}(t,\rho(\epsilon),\theta,\phi)}{\cos^3\! \rho(\epsilon)}\,.
\end{equation}

As a final note, let us add that our description of coherent states is conventional from a QFT perspective. However, the usual discussions of coherent states in the context of the AdS/CFT correspondence focus on the path integral preparation of these states by the introduction of sources in the boundary theory, \eg  \cite{BottaCantcheff:2015sav, Marolf:2017kvq,  BottaCantcheff:2019apr}. Ultimately, we are considering the same states as in those constructions.

%%%%%%%%%%%%%%%%%%%%%%%%
\section{\label{app:B} B. Holographic Complexity}

The details of the holographic complexity variation presented in section 3 are discussed below. The holographic complexity evaluated by the CA conjecture \eqref{defineCA} requires the addition of boundary contributions to have a well defined variational principle. The relevant action was worked out in \cite{Lehner:2016vdi} and, following the conventions of \cite{Chapman:2018dem}, reads: 
\begin{equation}
\begin{aligned}
I &= I_{\mt{bulk}}  + I_{\mt{GHY}} + I_{\mt{jt}} + I_{\kappa} + I_{\mt{ct}}  \\
&=  \frac1{16\pi\GN} \int d^{4} y \sqrt{-g} \bigg[{\cal R}+ \frac{6}{L^2}-
\frac1{2}  g^{\mu\nu}\nabla_\mu \Phi \nabla_\nu \Phi  \bigg]+ \frac{1}{8\pi \GN} \int_{\mt{regulator}}\!\!\!\!\!   d^{3}x\,\sqrt{|h|}\,K  \\
&\, + \frac{1}{8\pi \GN} \int_{\rm joints}\!\!\!\!\!  d^{2}x\,\sqrt{\sigma}\,a + \frac{1}{8\pi \GN} \int_{\del{\rm WDW}}\!\!\!\!\!  ds\,d^{2}\Omega\,\sqrt{\gamma}\,\kappa  + \frac{1}{8\pi \GN} \int_{\del{\rm WDW}}\!\!\!\!\! ds \,d^{2}\Omega\,\sqrt{\gamma}\, \Theta \log (\ell_{\rm ct} \Theta) 
\,. 
\end{aligned}
\label{eq:faction}
\end{equation}
$I_{\mt{bulk}}$ contains the Einstein-Hilbert action coupled to a (negative) cosmological constant and the bulk scalar field. There are several surface terms: $I_{\mt{GHY}}$ is the usual Gibbons-Hawking-York term defined in terms of the trace of the extrinsic curvature $K$ \cite{PhysRevLett.28.1082, PhysRevD.15.2752}. Here, it is evaluated at the asymptotic regulator surface, with induced measure $\sqrt{|h|}$, where the latter cuts off the WDW patch \cite{Carmi:2016wjl}. $I_{\mt{jt}}$ is the null joint term evaluated where the null boundaries of the WDW patch intersect the regulator surface. The measure $\sqrt{\sigma}$ stands for the induced metric on this joint and $a=\epsilon\log |k\cdot s|$ with $s_\alpha dx^\alpha$ and $k_\alpha dx^\alpha$, the outward-directed normals to both boundaries, respectively, and $\epsilon$ is a sign defined in \cite{Lehner:2016vdi}. Both $I_{\kappa} $ and $I_{\mt{ct}}$ involve integration over the null boundaries of the WDW patch and were described in the main text -- see eqs.~\reef{kappa} and \reef{ct33}. The first depends on the scalar $\kappa$ describing by how much the coordinate $s$ parameterising the null boundary direction fails to be an affine parameter. The second is the counterterm action introduced in \cite{Lehner:2016vdi} to ensure that the action is invariant under reparameterisations of the null boundary. It depends on an arbitrary scale $\ell_{\mt ct}$ and the expansion scalar of null generators $\Theta =  \partial_s \log \sqrt{\gamma}$, where $\sqrt{\gamma}$ is the induced measure on the null boundary. 

The holographic complexity variation considers the difference of the total action \eqref{eq:faction} evaluated on the perturbed configuration \eqref{metric} sourced by the scalar field \eqref{class2} in the deformed WDW patch  minus the same total action \eqref{eq:faction} evaluated on the undeformed WDW patch in the  AdS$_4$ vacuum \eqref{vac}. Hence, the second order action variation naturally splits into two contributions
\begin{equation}
  \delta I = I[g_0+\delta g,\,\Phi] - I[g_0] = \delta I_{\mt{WDW}}+ \delta I_{\delta\mt{WDW}} \, ,
\label{action_variation}
\end{equation}
where $ \delta I_{\mt{WDW}}$ is the variation due to the change in the background fields within the original WDW patch, while $\delta I_{\delta\text{WDW}}$ is the variation due to the change in the shape of the WDW patch (\ie the position of the boundary). 

Before analysing the different terms in \eqref{action_variation}, let us describe the null boundaries of the relevant WDW patches.
Given the spherical symmetry of the problem, the boundary of the perturbed WDW patch at order $\varepsilon^2$ is easily described in terms of the null condition
\be
\[ 1+ \varepsilon^2 (a_2 - 2 d_2) \]   dt^2 = (1- \varepsilon^2 a_2)  d\rho^2 \,.
\ee
If we choose the radius  $\rho$ as our parameter, one can perturbatively solve for the future($+$) and past($-$) directed null geodesics, $t = t_{\pm}(\rho) +  \delta t_{\pm}(\rho)$, originating on the AdS boundary $\rho = \pi/2$ at $t=0$. We obtain
\be \label{eq:WDWfull}
t_{\pm}(\rho) = \pm (\frac \pi 2 - \rho)\,  \qquad  \text{and} \qquad \delta t_{\pm}(\rho) = \mp \varepsilon^2\int^{\pi/2}_\rho  \(a_2(t_{\pm}(y),y)-d_2(t_{\pm}(y),y) \) dy \, . 
\ee
Given this description of the null boundaries, we can define outward-pointing normal one-forms and the corresponding null normal vectors as
\begin{equation}
\begin{split}
  k_\mu\, dx^\mu  &=  \pm dt  + d\rho - \varepsilon^2   \(a_2 -d_2 \) d\rho \, , \\
  k^\mu\, \del_\mu &=   \frac{\cos^2\rho}{L^2}  \[  \mp \del_t + \del_\rho   +   \varepsilon^2    \( \pm   \( a_2 - 2 d_2  \) \del_t  +  d_2  \del_\rho \) \]\equiv (k_0^\alpha +\delta k^\alpha)\del_\alpha \, .
\end{split}
\label{dk}
\end{equation}
Again, the upper (lower) sign choice corresponds to the upper (lower) boundary of the WDW patch. 
Notice that our choice of $s$, the null coordinate on the null WDW boundary satisfying the relation $\del_s \equiv k^{\mu}\del_{\mu}$,  is only affine at leading order in the perturbative expansion. For example, we see that $k_\mu dx^\mu$ is not a closed one-form at order $\veps^2$. Indeed, an explicit calculation of $k^{\mu} \nabla_{\mu} k_{\nu} = \kappa\,  k_{\nu}$ reveals
\begin{equation}
  \kappa =  \delta\kappa = \pm  \varepsilon^2 \frac{\cos^2\rho}{L^2} \del_t (a_2-d_2)\, . 
\label{kappadef}
\end{equation}

Now let us consider the various contributions to the variation \reef{action_variation} from all of the various terms in the action \reef{eq:faction}. First, let us discuss the contributions from $I_{\mt{GHY}}$ and $I_{\mt{jt}}$. Since both the AdS$_4$ vacuum \reef{vac} and our perturbed metric \eqref{metric} have no horizons in the bulk interior, \cite{foot02} the contributions from both $\delta I_{\mt{GHY}}$ and $\delta I_{\mt{jt}}$ only involve the behaviour of the metric perturbations close to the AdS boundary. Since the asymptotic decay of these perturbations is fast enough, both of these variations vanish in the limit where the AdS boundary regulator is taken to infinity. 

Hence, as claimed in the main text, we are left with $\delta I_{\mt{bulk}}+ \delta I_{\mt{ct}} + \delta I_{\kappa} $. 
It follows from eq.~\eqref{kappadef} that $\delta I_\kappa$ has an $\varepsilon^2$ contribution from integrating $\delta\kappa$ along the boundary of the unperturbed WDW patch 
\begin{equation} \label{eq:kappa}
  \delta I_{\kappa} =   \frac{1}{8\pi \GN}\int_{\del{\rm WDW}} \!\!\!\! ds \,d^{2}\Omega\,\sqrt{\gamma}\,  \delta\kappa\,.
\end{equation}

Next, we consider the counterterm contribution 
\begin{equation}
  \delta I_{\mt{ct}} =   \frac{1}{8\pi \GN} \int_{\partial\delta\mt{WDW}}\!\!\!\! ds\,d^{2}\Omega\,\sqrt{\gamma}\, (\Theta_0+\delta\Theta)\log \ell_{\mt{ct}}(\Theta_0+\delta\Theta) -  \frac{1}{8\pi \GN}\int_{\partial\mt{WDW}}\!\!\!\! ds\,d^{2}\Omega\,\sqrt{\gamma}\,\Theta_0\log \ell_{\mt{ct}}\Theta_0\,.
\label{dCT}
\end{equation}
Spherical symmetry guarantees the perturbed expansion $\delta\Theta$ is only due to the change in the tangent vectors \reef{dk} along the null boundaries 
\begin{equation}
  \delta \Theta = \delta k^\mu\,\partial_\mu \log \sqrt{\gamma}\,.
\end{equation}
Using the identities \cite{foot03}
\begin{equation}
\begin{split}
&   \int_{\partial\delta\mt{WDW}}\!\!\!\!\!\! ds\,d^{2}\Omega\,\sqrt{\gamma} \Theta_0  \log\ell_{\mt{ct}}\Theta_0 -   \int_{\partial\mt{WDW}}\!\!\!\!\!\! ds\,d^{2}\Omega\,\sqrt{\gamma} \Theta_0  \log\ell_{\mt{ct}}\Theta_0  = - \int_{\partial\mt{WDW}}\!\!\!\! ds\,d^{2}\Omega\left(\delta k^\mu\partial_\mu\log\sqrt{\gamma}\right)\log \ell_{\mt{ct}}\Theta_0\,, \\
  & (\Theta_0+\delta\Theta)\log \ell_{\mt{ct}}(\Theta_0+\delta\Theta) - \Theta_0\log\ell_{\mt{ct}}\Theta_0 = \delta k^\mu\partial_\mu\log\sqrt{\gamma} + \left(\delta k^\mu\partial_\mu\log\sqrt{\gamma}\right)\log \ell_{\mt{ct}}\Theta_0\,,
\end{split}
\end{equation} 
all the dependence on the arbitrary scale $\ell_{\mt{ct}}$ cancels, and the variation $\delta I_{\mt{ct}}$ reduces to the expression in eq.~\reef{hop99}, \ie
\begin{equation} \label{eq:ct}
   \delta I_{\mt{ct}} = \frac{1}{8\pi \GN}  \int_{\partial\mt{WDW}} \!\!\!\! ds\,d^{2}\Omega\,\sqrt{\gamma}\,\delta k^\mu\partial_\mu \log \sqrt{\gamma}\,.
\end{equation}

Finally, we consider the term $\delta I_{\mt{bulk}}$. Since we are considering the variation of the action evaluated on a solution of the equations of motion, the contribution to $\delta I_{\mt{WDW}}$ 
only gives rise to total derivatives which, using Stokes' theorem, translates into the boundary contributions
\begin{equation}
\begin{split}
  \delta I_{\mt{bulk,\,WDW}} &= \frac{1}{16\pi \GN}\int_{\mt{WDW}} \!\!d^4y\,\sqrt{-g_0}\nabla_\sigma\left(g_0^{\sigma\nu}\nabla^\mu\delta g_{\mu\nu} - \nabla^\sigma\delta g^\mu{}_\mu-\frac{1}{2}g_0^{\sigma\nu}\Phi \nabla_\nu\Phi \right) \\
  &= \frac{1}{16\pi \GN}\int_{\del\mt{WDW}} \!\!\!\!ds\, d^{2}\Omega\,\sqrt{\gamma}\, k_\sigma\left(g_0^{\sigma\nu}\nabla^\mu\delta g_{\mu\nu} - \nabla^\sigma\delta g^\mu{}_\mu-\frac{1}{2}g_0^{\sigma\nu}\Phi \nabla_\nu\Phi  \right)\, .
\end{split}
\end{equation}
Furthermore, at second order, the contribution to $\delta I_{\delta\mt{WDW}}$ can only originate from integrating the zeroth-order on-shell Lagrangian density over the additional spacetime volume enclosed by the perturbed WDW patch. This contribution can also be recast in the form of an integral over the boundary of the original WDW
\begin{equation}
\begin{split}
  \delta I_{\mt{bulk},\,\delta\mt{WDW}} &=  \frac{1}{16\pi \GN}\int_{\delta\mt{WDW}} \!\!\!\!\!
  d^4y\,\sqrt{-g_0}\bigg[{\cal R}(g_0)+ \frac{6}{L^2}\bigg] =   \frac{1}{8\pi \GN}  \frac{\varepsilon^2}{L^2}   \int_{\del \mt{WDW}} \!\!\!\!\! ds\,d^{2}\Omega\,\sqrt{\gamma}\,  \sin \rho \cos \rho  \(a_2 -d_2 \) \, .
\end{split}
\end{equation}
The second equality follows using the vacuum AdS$_4$  value of ${\cal R}(g_0) = -12/ L^2$, $\delta t_\pm(\rho)$ in eq.~\eqref{eq:WDWfull}, and rearranging the resulting order of integrations.
 
Combining the two bulk contributions, using integration by parts in some of the terms involving the metric perturbations and taking into account their fall-off at the AdS boundary, one is left with the bulk action variation
\be
\delta I_{\mt{bulk}} =  \delta I_{\mt{bulk,\,WDW}} + \delta I_{\mt{bulk},\,\delta\mt{WDW}} = \frac{ \varepsilon^2}{8\pi \GN}\int_{\del\mt{WDW}} \!\!\!\!ds\,d^{2}\Omega\,\sqrt{\gamma}  \( \mp  \frac{ \cos^2 \rho  }{L^2}\del_t (a_2-d_2) -  2 \frac{\cot \rho}{L^2} \, d_2  -  \frac{1}{8}  \del_s \( \Phi_{\mt{cl}}^2 \)  \)\,  . 
\ee
Once again the sign choice corresponds to terms on the upper (lower) boundary of the WDW patch. 

Substituting eq.~\eqref{kappadef} into eq.~\eqref{eq:kappa}, and $\delta k^\mu$ from eq.~\eqref{dk} into eq.~\eqref{eq:ct}, the sum $\delta I_{\kappa} + \delta I_{\mt{ct}}$ cancels the first two terms in $\delta I_{\mt{bulk}}$. Hence, all gravitational contributions to the holographic complexity at second order cancel out, and the total variation of the action is only sourced by the variation of the scalar field action
\be
\delta I  =   \delta I_{\mt{mat}} = - \frac{ \varepsilon^2}{64\pi \GN}\int_{\del\mt{WDW}} \!\!\!\!ds\,d^{2}\Omega\,\sqrt{\gamma} \,  \del_s\(\Phi_{\mt{cl}}^2\) \,,
\ee
matching the result in the main text, \ie eq.~\eqref{cavar}.

\end{document}